\documentclass[12pt]{article}
\usepackage[normalem]{ulem}
\usepackage[mathscr]{eucal}
\usepackage[version=3]{mhchem}
\usepackage{stmaryrd}

\usepackage{epsfig,latexsym,amsfonts,amsmath,amsthm,amssymb,amsbsy,multirow,slashed,wasysym,textcomp,dsfont,comment,mathtools,cancel,cite,datetime,appendix,BOONDOX-calo}
\usepackage{tocloft}
\usepackage{bbold}
\usepackage{tikz}
\usetikzlibrary{decorations.pathreplacing,calligraphy}
\usetikzlibrary{backgrounds}
\usetikzlibrary{patterns}
\usetikzlibrary{arrows.meta}
\tikzstyle{bag} = [align=center]
\usetikzlibrary{decorations.pathmorphing}
\def\bea{\begin{eqnarray}}
\def\eea{\end{eqnarray}}

\usepackage{hyperref}
\usepackage{subcaption}

\usepackage{array}
\usepackage{booktabs}

\newcommand{\badat}{\begin{alignedat}}
\newcommand{\eadat}{\end{alignedat}}
\newcommand\scalemath[2]{\scalebox{#1}{\mbox{\ensuremath{\displaystyle #2}}}}
\def\be{\begin{equation}}
\def\ee{\end{equation}}
\def\ba{\begin{aligned}}
\def\ea{\end{aligned}}

\usepackage{xcolor}

\newcommand{\pink}[1]{\textcolor{\pink}{#1}}

\definecolor{dblue}{rgb}{0.2,0.50,0.80}

\def\L{\mathcal{L}}

\def\bz{{\bar z}}
\def\bw{{\bar w}}

\def\L{\Lambda}

\def\bz{{\bar z}}
\def\bw{{\bar w}}

\def\pa{{\partial}}
\def\a{{\alpha}}

\def\d{{\delta}}
\def\o{{\omega}}
\def\b{{\beta}}

\def\m{{\mu}}

\def\l{{\lambda}}
\def\tl{{\tilde{\l}}}
\def\mg{{\mathscr G}}

\newcommand{\rvline}{\hspace*{-\arraycolsep}\vline\hspace*{-\arraycolsep}}

\DeclareFontFamily{OT1}{pzc}{}
\DeclareFontShape{OT1}{pzc}{m}{it}{<-> s * [1.10] pzcmi7t}{}
\DeclareMathAlphabet{\mathpzc}{OT1}{pzc}{m}{it}

\definecolor{vert}{rgb}{0.1367 0.543 0.1367}

\makeatletter
\def\blfootnote{\gdef\@thefnmark{}\@footnotetext}
\makeatother

\DeclareSymbolFont{extraup}{U}{zavm}{m}{n}
\DeclareMathSymbol{\varheart}{\mathalpha}{extraup}{86}
\DeclareMathSymbol{\vardiamond}{\mathalpha}{extraup}{87}

\numberwithin{equation}{section} % equation numbers follow sections

\begin{document}

 \begin{titlepage}
  \thispagestyle{empty}
  \begin{flushright}
%%%% report number
  \end{flushright}
  \bigskip
  \begin{center}

        \baselineskip=13pt {\LARGE {
        Celestial Quantum Error Correction II: \\[.5em]
        From Qudits to Celestial CFT 
       }}

      \vskip1cm 

   \centerline{ 
   {Alfredo Guevara}\footnote{aguevara@ias.edu} ${}^{\clubsuit,{}\spadesuit{},{}\vardiamond{}}$  
    and {Yangrui Hu}\footnote{{yangrui.hu1@uwaterloo.ca}} ${}^{\diamondsuit,{}\heartsuit{}}$ 
}

\bigskip\bigskip

\centerline{\em${}^\clubsuit$ 
\it School of Natural Sciences, Institute for Advanced Study, Princeton, NJ 08540 USA
}

\bigskip

\centerline{\em${}^\spadesuit$ 
\it Center for the Fundamental Laws of Nature, Harvard University, Cambridge, MA 02138 USA
}

\bigskip

\centerline{\em${}^\vardiamond$ 
\it Society of Fellows, Harvard University, Cambridge, MA 02138 USA
}

\bigskip

\centerline{\em${}^\diamondsuit$ 
\it Perimeter Institute for Theoretical Physics, Waterloo, ON N2L 2Y5, Canada}

\bigskip

\centerline{\em${}^\heartsuit$ 
\it Department of Physics and Astronomy, University of Waterloo, ON N2L 3G1, Canada}

\bigskip\bigskip

\end{center}

\begin{abstract}
 \noindent 
A holographic CFT description of asymptotically flat spacetimes inherits vacuum degeneracies and IR divergences from its gravitational dual. We devise a Quantum Error Correcting (QEC) framework to encode both effects as correctable fluctuations on the CFT dual. The framework is physically motivated by embedding a chain of qudits in the so-called Klein spacetime and then taking a continuum $N\to \infty$ limit. At finite $N$ the qudit chain 1) enjoys a discrete version of celestial symmetries and 2) supports a Gottesman-Kitaev-Preskill (GKP) code. The limit results in hard states with quantized BMS hair in the celestial torus forming the logical subspace, robust under errors induced by soft radiation. 
Technically, the construction leverages the recently studied $w_{1+\infty}$ hierarchy of soft currents and its realization from a sigma model in twistor space. 
  
\end{abstract}

\end{titlepage}

\tableofcontents

\section{Introduction}\label{sec:intro}

While a lot of the progress in the celestial/flat holography program has derived from reproducing IR physics in physical theories from a CFT description~\cite{Strominger:2017zoo}, a separate recent line of development aims to match certain structures, deriving from $w_{1+\infty}$ symmetries, from fundamental theories in twistor space~\cite{Adamo:2021bej,Adamo:2021lrv}. This poses an inherent paradox as twistor space theories do not inherently have IR divergences, i.e., IR modes could only emerge in an effective regime. To reconcile this requires finding a way to factorize a fundamental UV Hilbert space into hard and soft states in the presence of a cutoff scale, where the hard states would lead to IR safe observables.

How do degrees of freedom factorize and renormalize in gravitational theories? 
Over the last decade, impressive theoretical and experimental advances have been made in the topic of Quantum Error Correction (QEC) \cite{Roffe:2019srk}. They provide a link across diverse areas of physics, bridging many-body systems and quantum gravity. 

The connection between quantum information and holography has not been developed in the asymptotically flat case, but has been studied extensively in AdS/CFT. There, the application of error correction in quantum gravity can be traced back to an interplay between many-body entanglement and geometry. 
In this context, the Hilbert space is that of particular many-body systems whose operators and observables match those of a gravitational calculation \cite{Vidal:2007hda,Vidal:2008zz,Swingle:2009bg}. 
Advancing along this trajectory, a proposal for characterizing features in AdS/CFT using quantum error-correcting codes (QECC) was made in~\cite{Almheiri:2014lwa}. 
The basic setup of holographic QECC consists of approximately isometric maps
\begin{equation}
    V:~~ \mathcal{H}_{bulk} ~\longrightarrow~ \mathcal{H}_{boundary}~~,~~ |V|\psi\rangle|~\approx~||\psi\rangle| 
    \label{eq:mpas}
\end{equation}
that embed a bulk (logical) state into certain physical CFT states living on the AdS boundary.  
A particularly nice way of realizing the quantum code involving the isometry $V$ is by tiling a version of AdS spacetime with tensor networks, as illustrated in \cite{Pastawski:2015qua}. The output of the networks is given by qubits representing CFT operators in the dual theory. The number of external qubits increases (`renormalizes') asymptotically with the radial direction $R$, and the boundary theory emerges for $N \propto R \to \infty$ qubits. Because of this, in these codes, the emergent radial direction is a renormalization scale and measures how well boundary representations of bulk states are protected from errors. The map \eqref{eq:mpas} is nothing but a map from Hilbert spaces at finite $R$ to $R\to \infty$.

Is it feasible to extend analogous concepts to the flat (celestial) holography? Based on the above, it appears what we need to find is a type of many-body system that renormalizes $N\to \infty$ as we approach the asymptotically flat boundary. The system has to be directly tied to that topology, either a celestial sphere or a celestial torus where the CFT lives.
What symmetry should it possess? The asymptotically flat gravity features spontaneous (BMS-)symmetry-breaking with Goldstone modes (soft gravitons) responsible for infrared (IR) divergences~\cite{He:2014laa,Himwich:2020rro}. Such features are indeed present very generically in many-body systems and their continuum limit. However, a very particular extension of this symmetry arises on the CFT side: There exists an infinite tower of soft currents~\cite{Guevara:2019ypd} corresponding to soft gravitons that form the $w_{1+\infty}$ chiral symmetry algebra~\cite{Guevara:2021abz,Strominger:2021mtt}. Coincidentally, this tower admits a description in terms of theories living in twistor space, which, following the ideas of Penrose, we would like to interpret as carrying the fundamental degrees of freedom of gravity.

We will take this symmetry and its twistorial realization as a guideline to propose a quantum system where soft and hard modes factorize in an asymptotically flat topology. This system does this by storing and correcting quantum information.

In~\cite{Guevara:2023tnf}, we began 
our exploration by pinpointing that the same algebraic structure emerges from 1) a noncommutative version of Klein spacetime (generally the hyperkähler spaces described by twistor theories) and 2) the Gottesman-Kitaev-Preskill (GKP) code applied to a qubit system~\cite{Gottesman:2000di}. This leads to a very basic toy model for celestial CFT (CCFT) with finite quantum degrees of freedom. In this toy model, the code subspace is the Hilbert space of a two-qubit system protected against soft spacetime fluctuations. The aim of this work is to promote the toy model to a full-fledged CFT that manifests the above features from a QECC perspective. In close analogy with AdS/CFT, the guiding principle for implementing the code is the topology of the asymptotically flat spacetimes. In particular, we exploit the fact that celestial CFTs are suitable in $(2,2)$ signature, the so-called Kleinian spacetimes~\cite{Atanasov:2021oyu}. This signature is also natural for the construction of twistor spaces, which exhibit a dual description of the celestial CFTs. Using this machinery, we aim to
\begin{enumerate}
    \item show how the IR data is factored out, whereas the hard data is protected, i.e., free of IR divergence even in the presence of soft modes;
    \item determine what kind of errors can be reversed for asymptotic holographic states and show how they relate to standard memory;
    \item hint towards a bulk realization in terms of many-body states such as a spin chain, with the condition that it converges to the CFT.\,\footnote{Let us elaborate: One could envision the topology of asymptotically flat Kleinian spacetimes, namely a celestial torus, allowing for insertion of a chain qudits along one of its cycles. Consider such states in the torus at fixed radial distance $R$, as shown in Figure~\ref{fig:torus}. As we push $R\to \infty$, the chain is renormalized and it flows to a boundary CFT. The QECC can then be represented as an isometric map
\begin{equation}
    V:~~ \mathcal{H}_{R_0} ~~\longrightarrow~~ \mathcal{H}_{R\to \infty}
\end{equation}
where each Hilbert space is a tensor product $\mathcal{H}^{\otimes N}$ of $N\propto R^2$ qudits. This picture is one of our guiding motivations, and even though not strictly necessary for the construction, we find it useful to keep in mind. Furthermore, it closely mimics the AdS scenario \cite{Pastawski:2015qua}.}
\end{enumerate}

\subsection{Outline of applications}
Let us summarize part of the outcome of the construction by giving a comparison with the more traditional approach to IR effects.
Measurement of `holographic' gravitational modes is performed on the celestial sphere at fixed angles, see e.g. \cite{Himwich:2020rro}. For instance, soft gravitons can emerge in a detector as Bremsstrahlung, a.k.a. Weinberg's soft theorem. This involves measuring the correlation function
\begin{equation}
    \langle H(z)\, {\cal O}_1\ldots {\cal O}_n\rangle ~=~ \kappa \sum_i\frac{[\lambda \lambda_i]}{z-z_i} \langle  {\cal O}_1\ldots {\cal O}_n\rangle  \label{eq:xq2}
\end{equation}
(see section \ref{sec:celestialcode} for notation) where $H(z)$ is a soft graviton and $z\in \mathbb{CP}^1$ can be interpreted as a fixed angle in the celestial sphere.

\begin{figure}[t]
    \centering
    \includegraphics[width=.5\textwidth]{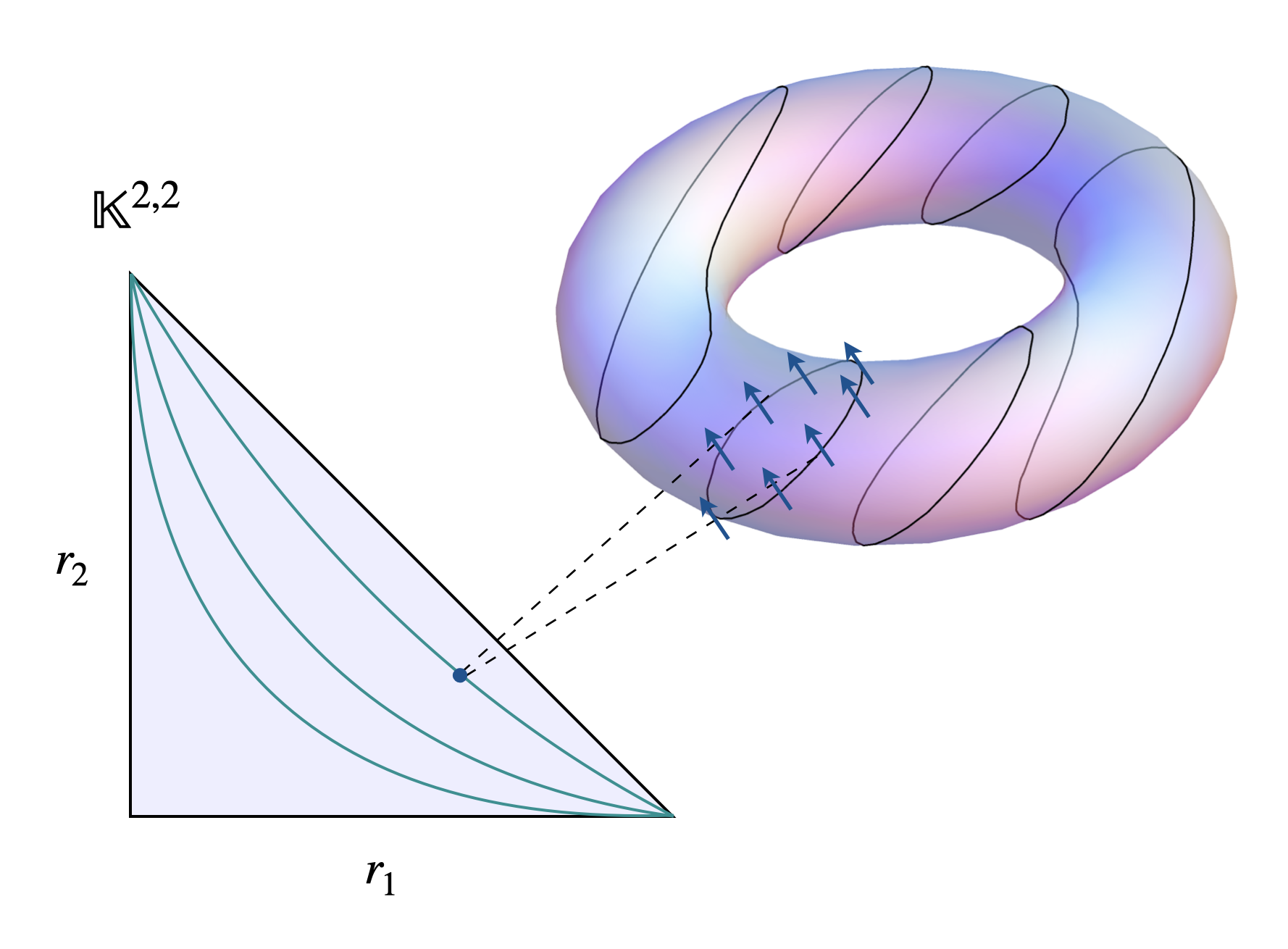}
    \caption{The left shows a toric Penrose diagram for the flat Klein spacetime $\mathbb{K}^{2,2}$ with metric $ds^2=dr_1^2+r_1^2d\phi^2-dr_2^2-r_2^2d\psi^2$. 
    As demonstrated, a torus is fibered over each point in the diagram. Green curves represent hyperbolae of constant torus area $r_1r_2$, while black contours on the torus correspond to $x^+=\phi-\psi$. As explained in section~\ref{sec:qudit}, we insert $N$ qudits along the $x^+$ cycle, where $N$ is proportional to the radial distance $R^2=r_1^2-r^2_2$. Hence as approaching the null boundary, $R^2\to\infty$ and $N\to\infty$, the $N$-qudit system flows towards a CFT as its continuum limit. }
    \label{fig:torus}
\end{figure}

By constructing a multi-qudit system (with $N$ degrees of freedom per site), and then taking $N\to\infty$ we are led to a non-local measurement in the boundary CFT, taking the schematic form
\begin{equation}
    \langle \exp\left(N\int_{\cal R} dz z^{k-1/2} H(z )  \right ) {\cal O}_1 \ldots {\cal O}_n\rangle ~\sim~ e^{2\pi i \kappa}\, \langle {\cal O}_1 \ldots {\cal O}_n\rangle ~~.
    \label{eq:xq2-exp}
\end{equation}
The exponential operator inserted here turns out to be a stabilizer of the code, namely a diagnosis of the `error' $\kappa$ induced by other states. In the CFT, it corresponds to measuring the graviton state as integrated along a null direction in the celestial torus (see Fig. \ref{fig:torus}), which can be identified with a certain light-ray operator~\cite{Sharma:2021gcz,Guevara:2021tvr,Hu:2022syq,Hu:2022txx,Hu:2023geb}. The result of the measurement depends on the region ${\cal R}$ and encapsulates the insertion of soft gravitons in such region. Furthermore, our proposal is consistent with the full tower of $w_{1+\infty}$ Ward identities, as can be guessed from the exponential structure of \eqref{eq:xq2-exp}. In particular, it connects naturally to the explicit realization of this symmetry in twistor space as described in~\cite{Adamo:2021bej,Adamo:2021lrv}.

The outline of the paper is as follows. In section~\ref{sec:GKP} we revisit the quantum error-correcting code based on the GKP construction~\cite{Gottesman:2000di}, see also~\cite{Calderbank:1996hm,Gottesman:1996rt}. Section~\ref{sec:qudit} is devoted to exploiting the idea of GKP in celestial holography, where we promote the two-qubit toy model proposed in~\cite{Guevara:2023tnf} to an $N$-qudit system and study its symmetries. 
The main result, presented in section~\ref{sec:celestialcode}, is a code that protects only quantized supertranslation charges from soft fluctuations. This sheds light on the long-standing problem of constructing quantized charges for gravitational states in an analogy of quantized QED charges living on a lattice~\cite{Nande:2017dba}.

\section{GKP codes revisited}\label{sec:GKP}

The main application of GKP encoding is to embed a small-dimensional (logical) Hilbert space into a large-dimensional (physical) Hilbert space \cite{Gottesman:2000di}. The physical system can be either finite or infinite, discrete or continuous. Our task in this section is two-fold. We first want to review and revisit the GKP construction in our context, applied to a single qudit. At the same time, we want to reinterpret the qudit as a toy model for encoding information in a gravitational system. 

The last statement requires some unpacking. In the context of asymptotically flat gravity, what is the physical system where the qudit shall be embedded? A pressing issue in the description of asymptotically flat spacetimes, via e.g. celestial holography, is the identification of an appropriate vacuum state incorporating long wavelength radiation~\cite{Melton:2023dee}. It has been proposed that holographic CFT states appear naturally dressed by a coherent cloud of IR radiation~\cite{Kapec:2017tkm,Himwich:2020rro,Arkani-Hamed:2020gyp}. We will attempt to shed light on this issue from a quantum information perspective: We seek a physical system where we can single out a unique vacuum embedded within coherent states. Turns out, the simplest system for such a task is the single harmonic oscillator. GKP codes precisely provide a way to incorporate spontaneous symmetry breaking into the oscillator. For our goals, this is a very basic toy model for celestial holography, while the more realistic model will be introduced in the next section based on the twistorial toolkit.

The basic idea is as follows.
Consider a single quantum harmonic oscillator, as given by a Hilbert space representing the commutation relations 
\begin{equation}
    \big[a,a^{\dagger}\big] ~=~ 1 ~~.
\end{equation}
Coherent states satisfying $a|\alpha\rangle=i\alpha  |\alpha \rangle$ are constructed as 
\begin{equation}\label{eq:glam}
   |\alpha\rangle ~=~ G_{\alpha} |0\rangle ~~,~~ G_{\alpha} =\exp{i\Big(\a\,a^{\dagger}+ \a^*a  \Big)} 
   ~~,
\end{equation}
where $G_{\alpha}$ is the usual displacement operator. Note that the oscillator Hamiltonian $\tilde{H}=a^{\dagger}a$ has $|\alpha = 0\rangle$ as its \textit{unique} vacuum. However, to obtain a toy model for holography, we will introduce a `symmetry-breaking pattern' associated with a different vacuum/Hamiltonian. For this, it is enough to fix a Hamiltonian that breaks the symplectic symmetries of phase space. We introduce `$XP$ coordinates'
\begin{equation}
    \mu_- ~=~ \sqrt{\frac{\tau}{2}}\,(a+a^{\dagger}) ~~,~~
    \mu_+ ~=~ -i\,\sqrt{\frac{\tau}{2}}\,(a-a^{\dagger}) ~~,
\end{equation}
which introduces an arbitrary modular parameter $\tau$. Thus, here we are considering a single `$\mu$-system' given by
\begin{equation}
     \big[\mu_-,\,\mu_+\big] ~=~ i\,\tau  ~~. \label{eq:comt}
\end{equation}
Note that in Klein space, these operators are both hermitian, hence the extra factor of $i$. The canonical transformations here include ${\rm SL}(2,\mathbb{R})$ symmetry $\mu_\alpha \to \Lambda_{\alpha}{}^{\beta}\, \mu_\beta$, with $\alpha,\beta=\pm$, together with shifts $\mu_{\pm} \to \mu_{\pm}+ c_{\pm}$. The latter will not be very relevant. In this example, symmetry breaking arises if we attempt to choose a `free theory' with a Hamiltonian of the form $H=\mu_+\mu_{-}+E_0$ which is \textit{not} the standard harmonic oscillator\,\footnote{Although it \textit{would be} in the Euclidean case! Since $\mu_\pm$ indeed play the role of $a,a^{\dagger}$ there. On the other hand, Hamiltonians of the type $XP$ are sometimes called inverted harmonic oscillators. See e.g. \cite{Parmentier:2023axg,Hadar:2022xag} for relevant discussions.}. Its vacuum is chosen as
\begin{equation}
    \mu_{+}|\tilde{0}\rangle~=~0~~  \textrm{or} ~~ \mu_{-}|\tilde{0}\rangle~=~0 \label{eq:choicev}
\end{equation}
which clearly does not respect  the canonical ${\rm SL}(2,\mathbb{R})$ symmetry. These are the ``$X$ and $P$ vacua" exchanged under symplectic rotations $X \to -P, P \to X $, e.g. a Fourier transform \cite{Gottesman:2000di}. 

Suppose we try to insist on a free vacuum that obeys \eqref{eq:choicev} simultaneously, therefore respecting ${\rm SL}(2,\mathbb{R})$. Even though this is not possible, we obtain an almost equivalent condition if we set a special value for $\tau$:\,\footnote{In the error correction literature, it is customary to take $N$ prime, e.g. \cite{Gottesman:1998hu}. This simplifies some aspects of the discussion but will be irrelevant for our purposes here, as $N\to \infty$.}
\begin{equation}\label{eq:discre}
    \tau ~=~ \frac{2\pi}{N}~~,~~ N\in 2\mathbb{Z}_+ ~~.
\end{equation}
Let us see why this is the case. 
We introduce the unitary operators
\begin{equation}\label{eq:stabilizer}
    g_\pm ~:=~ e^{i\mu_{\pm}}\,\quad \textrm{and}\,\quad S_{\pm}:~=~g_{\pm}^N \,
\end{equation}
which satisfy $g_+ g_- = e^{i\tau}g_- g_+$. 
Unlike $\{\mu_{\pm}\}$, the exponentiated operators $S_{\pm}$, called \textit{stabilizers}, commute and can be diagonalized simultaneously. We can think of them as abelian directions in phase space \cite{Gottesman:2000di}, thus we can use them to quotient such space into a lattice. This is done by picking any reference state $|r\rangle$ and defining
\begin{equation}\label{eq:quots}
    |\tilde{0}\rangle ~=~ \sum_{k,j\in \mathbb{Z}} S_{+}^k S_{-}^j |r\rangle   ~~.
\end{equation}
So instead of imposing \eqref{eq:choicev} simultaneously, we find a weaker condition
\begin{equation}\label{eq:vaccon}
    S_{\pm} |\tilde{0}\rangle ~=~ |\tilde{0}\rangle ~~.
\end{equation}
Now, does this vacuum choice preserve ${\rm SL}(2,\mathbb{R})$? Since $N$ is an integer, it turns out this choice only preserves a ${\rm SL}(2,\mathbb{Z})$ subgroup of symplectic transformations. Namely, $\mu_\alpha \to \Lambda_{\alpha}{}^{\beta}\, \mu_\beta$ leads to
\begin{equation}
    S_{\pm} ~\to~ S_+^{\Lambda_{\pm}^{+}}\,S_{-}^{\Lambda_{\pm}^-} ~~,\label{eq:2szt}
\end{equation}
and it is straightforward to check that \eqref{eq:quots} is invariant only if $\Lambda_{\alpha}{}^{\beta}$ are integers satisfying
\begin{equation}
    \Lambda_{+}{}^{+}\, \Lambda_{-}{}^{-} -  \Lambda_{-}{}^{+} \,\Lambda_{+}{}^{-} ~=~ 1 ~~.
\end{equation}
Hence the breaking
\begin{equation}
  {\rm Sp}(2,\mathbb{R})~\cong~{\rm SL}(2,\mathbb{\mathbb{R}})~\to~ {\rm SL}(2,\mathbb{Z})  
\end{equation}
corresponds to the modular group of the two-dimensional lattice generated by $S_{\pm}$. Indeed, we now show that this is a quantized phase space, with unit cells having area $\tau=2\pi/N$.

\subsection{Code Subspace}

The quantized phase space is a subspace of the operator algebra of the harmonic oscillator, called logical operators in the literature. They act on a subspace of the Hilbert space of the harmonic oscillator called \textit{code subspace}. It will be defined by the stabilizer conditions $S_{\pm}=1$ which are invariant under ${\rm SL}(2,\mathbb{Z})$. Note that $|\tilde{0}\rangle$ indeed lives in the code subspace as \eqref{eq:vaccon}. What other states satisfy this condition?

To find the full quantized phase space, consider the (overcomplete) set of states that can be obtained from $|\tilde{0}\rangle$ by the displacement operator \eqref{eq:glam}, which we denote as
\begin{equation}\label{eq:stateop}
     |\lambda\rangle ~=~ G_\lambda |\tilde{0}\rangle~~,~~ (\l_+,\l_-)~\in~ \mathbb{R}^2 ~~.
\end{equation}
Here we have written the coherent value as
\begin{equation}
    \alpha~=~\sqrt{\frac{\tau}{2}}\,(\lambda^{-} + i \lambda^{+})
\end{equation}
so that we can write
\begin{equation}\label{equ:gate}
    G_{\lambda}  ~=~ \exp{i\,\Big(\lambda^+\mu_+ + \lambda^-\mu_-\Big)}~=~e^{\pi i\frac{\lambda^+ \lambda^-}{N}}g_+^{\lambda^+} g_-^{\lambda^-} ~~.
\end{equation}
The displacements are unitary transformations that satisfy the Weyl algebra
\begin{equation}\label{eq:weylal}
    G_{\lambda_1} G_{\lambda_2} ~=~ e^{i\frac{\tau}{2} [\lambda_1 \lambda_2]} G_{\lambda_1 +\lambda_2}~=~e^{i\tau [\lambda_1 \lambda_2]} G_{\lambda_2} G_{\lambda_1} ~~,~~ [\lambda_1 \lambda_2] ~\equiv~ \lambda_1^+ \lambda_2^- - \lambda_1^- \lambda_2^+~~.
\end{equation}
The states we have isolated have a natural interpretation in CFT since \eqref{eq:stateop} is essentially the state-operator correspondence. Indeed, \eqref{eq:weylal} can be formulated as an operator map:
\begin{equation}
     G_{\lambda_1}|\lambda_2\rangle ~=~ e^{i\frac{\tau}{2}[\lambda_1 \lambda_2]}|\lambda_1 + \lambda_2\rangle ~~.
\end{equation}
Note that the two displacements defining the phase space are
\begin{equation}\label{eq:spm}
    S_+ ~=~ G_{(0,N)}~~,~~ S_- ~=~ G_{(N,0)}~~.
\end{equation}
Now, we say that $G_{\lambda}$ is in the quantized phase space (i.e. logical) if and only if $|\lambda\rangle$ is in the code space, namely
\begin{equation}
    S_{\pm}|\lambda\rangle ~=~|\lambda \rangle ~~.
\end{equation}
Since $|\tilde{0}\rangle$ is in the code, all we need for this is that $G_{\lambda}$ preserves the eigenvalues of $S_{\pm}$. From \eqref{eq:weylal}-\eqref{eq:spm} we find
\begin{equation}
    [G_{\lambda}, S_{\pm}] ~=~ 0 ~\Longrightarrow~ (\l_+,\l_-)~\in~ \mathbb{Z}_N \times \mathbb{Z}_N~~.
\end{equation}
The restriction to $\mathbb{Z}_N$ comes from the identification $\lambda_{\pm} \sim \lambda_{\pm}+N$ when acting on code states\,\footnote{Up to a possible minus sign in $G_{\lambda}$ which is irrelevant when acting on states. Formally, these operators live in a $\mathbb{Z}_2$ orbifold. In the CFT, the minus sign emerges after changing Poincaré patches in the Lorentzian cylinder, see~\cite{Kravchuk:2018htv}.}. Thus, these operators form a closed algebra of dimension $N^2$, namely the logical algebra, with $S_{\pm}$ lying in the center. Of course, we have already encountered the case $N=2$ in \cite{Guevara:2023tnf} (corresponding to $U(2)$), where $G_{(1,0)}=\sigma_{1}, G_{(0,1)}=\sigma_{3}$. 

The phase space is composed of the integer points $(\lambda_+,\lambda_-)$ playing the role of $X,P$ coordinates. Each $XP$ cell has volume $\tau=2\pi/N$ as read from \eqref{eq:comt}.
Due to the uncertainty principle of quantum mechanics, the phase space is of dimension $N^2$ but the Hilbert space ${\cal H}_{N}$ is $N$ dimensional. 

Although it is not necessary in what follows, sometimes it is convenient to write a basis for ${\cal H}_{N}$, 
since this will realize the $N^2$ operators as unitary transformations $U(N)$. More precisely, we can choose two different vacua
\begin{equation}\label{equ:GKPvacuum}
    | 0 \rangle_{\pm} ~=~\sum_{\lambda^{\pm}\in \mathbb{Z}_N\times \mathbb{Z}_N} e^{\pm \frac{\pi i 
\lambda^+ \lambda^- }{N}} |\lambda\rangle 
 ~=~ \sum_{\lambda^{\pm}\in \mathbb{Z}_N\times \mathbb{Z}_N} g_{\pm}^{\lambda^\pm } g_{\mp}^{\lambda^\mp }|\tilde{0}\rangle 
\end{equation}
which satisfy $g_{\pm}|0\rangle_{\pm} = |0\rangle_{\pm}$. We can then construct qudits by acting with $g_{\mp}$, namely\,\footnote{In this context $g_{\pm}$ are called generalized Pauli gates, usually denoted by $X$ and $Z$ in the literature.}
\begin{align}
|n\rangle_{\pm} ~=~ g_{\mp}^n  |0\rangle_{\pm}~~,~~ n=0,\ldots,N-1~~.
\end{align}

\subsection{Errors}\label{sec:qudit-error}

The quantized phase space, and the code states, are robust under small unitary fluctuations. These errors are represented by displacements $G_{\alpha}$ with $|\alpha|\ll 1$ and will be realized by soft radiation on the CFT side.

To describe errors, it is useful to first understand the full phase space. Recall that $\mu_+,\mu_-$ play the role of $X,P$ operators. We can label their spectrum by real eigenvalues $s_+$ and $s_-$ of stabilizers.  Since $S_{\pm}$ are simultaneously diagonalizable we can use $s_{\pm}$ as quantum numbers, more precisely
\begin{equation}
    S_{\pm}|\psi \rangle_s  ~=~ e^{iNs_{\pm}} |\psi\rangle_s ~~,
\end{equation}
where we denote $s=(s_+,s_-)$. Here $s_{\pm}$ is usually referred to as \textit{error syndrome}. They will play the role of Goldstone modes in CFT, in analogy with the gauge theory case \cite{Nande:2017dba}. Note that they live in the torus $S^1\times S^1$, namely
\begin{equation}\label{eq:toruss}
    s_{\pm} ~\sim~  s_{\pm} ~+~ \frac{2\pi}{N} ~~.
\end{equation}

The projection from the physical Hilbert space to the logical subspace is defined by the stabilizers: the code is spanned by stabilizer eigenstates with eigenvalue one, or $(s_+,s_-)=(0,0)$. We refer to the other eigenspaces $(s_+,s_-)\neq(0,0)$ in the torus as \textit{errors}. 

Soft errors acting on our code space are small fluctuations in $X,P$ eigenvalues and can be expressed as follows
\begin{equation}
    E_{\hat{\epsilon}} ~=~ \exp{i\Big(\hat{\epsilon}^+\mu_+ + \hat{\epsilon}^-\mu_-\Big)} 
   ~~.
\end{equation}
They can be diagnosed by evaluating the stabilizer. Using $S_{\pm}\, E_{\hat{\epsilon}}= e^{2\pi i\hat{\epsilon}_{\pm}}E_{\hat{\epsilon}}\,S_{\pm}$ we get
\begin{equation}
    E_{\hat{\epsilon}}\,|\psi\rangle_s ~=~ |\psi'\rangle_{s'} ~~,
\end{equation}
where the syndrome (i.e. ``Goldstone mode'') shifts as
\begin{equation}
    s'_\pm ~=~ s_\pm ~+~\frac{2\pi}{N}\,\hat{\epsilon}_{\pm} ~~.
\end{equation}
In particular, we note that in these codes, errors can be diagnosed when 
\begin{equation}
    |\hat{\epsilon}_{\pm}|~<~\frac{1}{2}~~,
    \label{equ:smallerror}
\end{equation}
and be reversed by acting $E^{\dagger}_{\hat{\epsilon}}=E_{-\hat{\epsilon}}$.  Clearly, an error $\hat{\epsilon}_{\pm}=1$ (or any integer) turns into a logical operation and cannot be detected. 
This is the implication of the torus identification \eqref{eq:toruss}. 

\begin{figure}[t]
\centering
\begin{tikzpicture}
        \draw[white, fill= cyan!10!white ] (-1.5/2,1.5/2) --  (1.5/2,1.5/2) -- (1.5/2,-1.5/2) -- (-1.5/2,-1.5/2) -- (-1.5/2,1.5/2);
        \filldraw[thick,fill=black] (0,0) circle (0.07);
        \filldraw[thick,fill=black] (1.5,0) circle (0.07);
        \filldraw[thick,fill=black] (3,0) circle (0.07);
        \filldraw[thick,fill=black] (-1.5,0) circle (0.07);
        \filldraw[thick,fill=black] (-3,0) circle (0.07);
        \filldraw[thick,fill=black] (0,-3) circle (0.07);
        \filldraw[thick,fill=black] (1.5,-3) circle (0.07);
        \filldraw[thick,fill=black] (3,-3) circle (0.07);
        \filldraw[thick,fill=black] (-1.5,-3) circle (0.07);
        \filldraw[thick,fill=black] (-3,-3) circle (0.07);
        \filldraw[thick,fill=black] (0,1.5) circle (0.07);
        \filldraw[thick,fill=black] (1.5,1.5) circle (0.07);
        \filldraw[thick,fill=black] (3,1.5) circle (0.07);
        \filldraw[thick,fill=black] (-1.5,1.5) circle (0.07);
        \filldraw[thick,fill=black] (-3,1.5) circle (0.07);
        \draw[thick] (-1.5/2,1.5/2) --  (1.5/2,1.5/2) -- (1.5/2,-1.5/2) -- (-1.5/2,-1.5/2) -- (-1.5/2,1.5/2);
        \filldraw[thick,fill=black] (0,-1.5) circle (0.07);
        \filldraw[thick,fill=black] (1.5,-1.5) circle (0.07);
        \filldraw[thick,fill=black] (3,-1.5) circle (0.07);
        \filldraw[thick,fill=black] (-1.5,-1.5) circle (0.07);
        \filldraw[thick,fill=black] (-3,-1.5) circle (0.07);
        \draw[thick,red] [decorate,decoration = {brace,mirror}] (0,-1.65) --  (1.5,-1.65);
        \draw[red](1.5/2,-1.7) node[below]{$\sqrt{\frac{2\pi}{N}}$};
        \draw[thick,cyan,<->] (0.1,0) -- (1.5/2,0);
        \draw[thick,cyan,<->] (0,0.1) -- (0,1.5/2);
        \draw[red](0.1+0.65/2,0) node[below]{$\hat{\epsilon}_{+}$};
        \draw[red](0,0.1+0.65/2) node[left]{$\hat{\epsilon}_{-}$};
        \draw(3.5,2) node[right]{\large $\l_{\pm}$};
\end{tikzpicture}
\caption{The quantized phase space, also called stabilizer lattice, is determined by $\lambda_{\pm} \in \mathbb{Z}_N$. The cyan-shaded square denotes the range of correctable errors.}
\label{fig:lattice}
\end{figure}
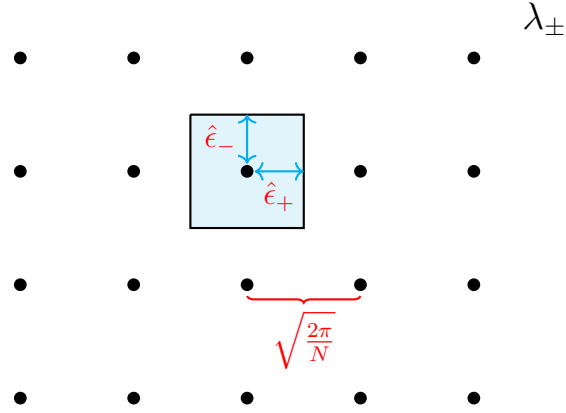

Recall that in the phase space the size of the unit cell is $\Delta \mu_+ \Delta\mu_{-} = \tau$, according to the uncertainty principle. In these units the correctable errors are fluctuations of magnitude $\sqrt{\tau}/2=\sqrt{\frac{\pi}{2N}}$, see Figure~\ref{fig:lattice}\,\footnote{{Note that the area of the unit cell in phase space being $\tau$ does not yield $\Delta\mu_+=\Delta\mu_-$. There is an additional tunable parameter for the relative uncertainty (denoted by $\alpha$ in~\cite{Gottesman:2000di}). Here we fix $\Delta\mu_+=\Delta\mu_-$ to respect ${\rm SL}(2,\mathbb{R})$ symmetry.}\label{ft:ex-sym-alpha}}. In particular at large $N$ we find 
\begin{equation}
    \Delta \hat{\epsilon}_{\pm}~=~\sqrt{\frac{\pi}{2N}} ~\ll~ 1 ~~.
\end{equation}
Interpreting $\sqrt{\frac{\pi}{2N}}$ as an IR cutoff, we see in our toy model that errors will correspond to an IR effect, namely soft radiation in the celestial context.

\section{Klein Space and Qudits}\label{sec:qudit}

In this section and the next section, we will apply the preceding discussion to celestial holography. To do so, let us now revisit the structure of $(2,2)$ spacetimes and its associated twistor fields. In the earlier work~\cite{Guevara:2023tnf} we mostly focused on flat Klein spacetime, namely 
\begin{equation}
    ds^2~=~ dz_1d\bar{z}_1 +dz_2d\bar{z_2} ~~, 
\label{eq:line}
\end{equation}
and analyzed a very simple toy model: A 2-qubit system obtained from its twistor field. More precisely, one can consider a sigma model with a twistor space target, parametrized as 
\begin{equation}
    \label{eq:wqew}
\mu_{\alpha}(z)~=~\underbrace{\sum_{k=-1/2}^{1/2}\frac{\mu_{\alpha}^{(k)}}{z^{k+1/2}}}_{\rm global}~+~\underbrace{\sum_{|k|>\frac{1}{2}}\frac{\mu_{\alpha}^{(k)}}{z^{k+1/2}}}_{\text{soft hair}} ~~,~~ \alpha=\pm~~,
\end{equation}
where $z\in \mathbb{RP}^1$ is a homogeneous real coordinate that can be associated with a (celestial) torus direction of Klein space \eqref{eq:line}, via Penrose's construction. Quantization of $\mu^{(\pm1/2)}_{\alpha}$ leads to a 2-qubit space equivalent to the quantum geometry of \eqref{eq:line}. This is where the error-correcting code acts.

We want to consider gravitational perturbations of \eqref{eq:line}. For this, we will study `Virasoro descendants' emerging as non-global modes 
\begin{equation}
    \mu_{\alpha}^{(k)}~~,~~|k|>1/2 ~~.
\end{equation}
They turn out to be associated with infinitely many `qudits'. Qudits are states in an $N$-dimensional Hilbert space $\mathcal{H}_N$, which are acted on by certain generalized versions of the Pauli gates.\,\footnote{We refer to~\cite{Guevara:2023tnf} for a somewhat self contained introduction.} These lie at the core of the error correction algorithm. In this section, we will first motivate the emergence of the qudit Hilbert space and then introduce the error correction algorithm for such. The latter mostly relies on the GKP codes~\cite{Gottesman:2000di} reviewed in the previous section. These are engineered for errors whose amplitude is modulated i.e. they are soft in a suitable sense.

\subsection{From Qubits to Qudits}\label{subsec:qudit}

We will argue that qudits emerge naturally from a certain discretization of twistor theory. 
First, we consider twistor field $\mu_{\a}(z)$ \eqref{eq:wqew} with coordinate $z\in \mathbb{RP}^1$. This line can be mapped to a circle $S^1$ by a conformal transformation. We will consider a non-hermitian field $\tilde{\mu}$ on this $S^1$,
\begin{equation}
    \tilde{\mu}_{\a}(x^+)~=~ \sum_{k}\mu_{\alpha}^{(k)}\,e^{ikx^{+}} ~~\,,\quad x^{+} \in [0,2\pi) ~~.
    \label{eq:2modes} 
\end{equation}
Under generic assumptions outlined in appendix~\ref{appen:modes},\footnote{The sigma model we have in mind requires the conformal field $\mu_{\a}(z)$ to be defined on a region of $\mathbb{CP}^1$, instead of just a real line~\cite{Mason:2022hly}.} the fields $\mu_{\a}(z)$ and $\tilde{\mu}_{\a}(x^+)$ are related by a simple analytic continuation and we will regard them as equivalent. We regard this $S^1$ as half the cycle of the Klein space celestial torus as follows. 
The condition of $h=1/2$ in the celestial torus can be stated as 
\begin{equation}
    \tilde{\mu}_\alpha(x^+ + 2\pi)  ~=~ - \tilde{\mu}_\alpha(x^+) ~~
    \label{eq:tcn}
\end{equation}
(we note that the shift $x^+\to x^+ + 2\pi$ corresponds to the action of $T$ for $\mu_{\a}(z)$ in \eqref{eq:wqew}
\begin{equation}
    T\,:~~ \mu_{\a}^{(k)}~\mapsto~ -\,\mu_{\a}^{(k)} ~~,
    \label{eq:tttp}
\end{equation}
which is equivalent to the time reversal and spatial reflection $(z_i,\bar{z}_i) \to -  (z_i,\bar{z}_i)$ as discussed in~\cite{Guevara:2023tnf}). 
The condition \eqref{eq:tcn} restricts \eqref{eq:2modes} to the expansion
\begin{equation}
\tilde{\mu}_\alpha(x^+) ~=~ \sum_{k\in\frac{1}{2}+\mathbb{Z}}\,\tilde{\mu}_\alpha^{(k)}\,e^{ikx^+}~~. \label{eq:mufl}
\end{equation}
This is indeed the most general form of the incidence relation in real twistor space, including conformal descendants as in e.g. \cite{Adamo:2021bej}. Note that the modes $\tilde{\mu}_\alpha^{(k)}$ are parity odd under $T$
and thus generalize the notion of $z,\bar{z}$  coordinates in \eqref{eq:line}: They are also Darboux coordinates but for twistor space.

The emergence of $N$-dimensional Hilbert spaces follows from truncating the series, after which we can take $N\to \infty$. This occurs as follows: insert operators at $N$ sites uniformly along the torus, located at angles
\begin{equation}
    x^+_j ~=~ \frac{2\pi j}{N} ~~,~~ j=0,\ldots,N-1~~.
\end{equation}
Each insertion introduces a Hilbert space $\mathcal{H}$ defined by the algebra
\begin{equation}
    \Big[\tilde{\mu}_{-}(x_{j}^{+}),\,\tilde{\mu}_{+}(x_{k}^{+})\Big]~=~i\tau\,\delta_{jk} ~~.
    \label{eq:rr1p}
\end{equation}
This is not only a canonical quantization of the circle field, but actually the precise symplectic form of twistor space.\,\footnote{The discussion can be extended to the more generic case $\Big[\tilde{\mu}_{\alpha}(x_{k}^{+}),\,\tilde{\mu}_{\beta}(x_{l}^{+})\Big] = \Omega^{kl}_{\alpha \beta}$. As long as the symplectic form $\Omega^{kl}_{\alpha \beta}$ is non-degenerate, the Darboux frame \eqref{eq:rr1p} can always be attained locally in twistor space.} We regard $\tau$ as a central term. The system has a representation in $\mathcal{H}^{\otimes N}$.\,
Since we only insert operators at angles $x_j^+$, we can set $\tilde\mu^{(k)}_\alpha= 0$ for $|k|>N/2$, leading to
\begin{equation}
\tilde{\mu}_\alpha(x^+) ~~\to~~ \sum_{k=-\frac{N-1}{2}}^{\frac{N-1}{2}}\,\tilde{\mu}_\alpha^{(k)}\,e^{ikx^+} ~~.
\label{equ:truncated-mode-exp}
\end{equation}
The commutation relations of the modes follow from \eqref{eq:rr1p} and become
\begin{equation}\label{equ:tildemukalg}
\Big[\tilde{\mu}_{-}^{(k)},\,\tilde{\mu}_{+}^{(j)}\Big]~=~i\tilde{\tau}\,\delta^{j+k}~~,~~ k,j=-\frac{N-1}{2},\ldots,\frac{N-1}{2}
\end{equation}
with
\begin{equation}
    \tilde{\tau}~=~\tau/N ~~. 
    \label{equ:tau-renormN}
\end{equation}
This system is the generalization of the 2-qubit system of~\cite{Guevara:2023tnf}. Indeed, it is easier understood as $N/2$ decoupled systems. 

The physical character of these operators as a certain many-body system/spin-chain shall be addressed in the next section. In this section, we focus on their quantum information aspect. For now, let us simply note that the scaling of $\tilde{\tau}$ as $1/N$ is a consequence of inserting $N$ modes along the $x^+$ cycle. The large $N$ limit, $N\to\infty$ is anticipated as the continuum limit for the spin chain. 
This is 
consistent with our expectations from renormalization $\tilde{\tau}=\tau/R^2$~\cite{Guevara:2023tnf}, where $R^2$ is the radial distance introduced from (\ref{eq:line}). 
Namely the number of qudits increases with the area of the torus, $N \propto R^2$, as we approach the boundary $R\to \infty$.\,\footnote{Note that the radius of the $x^+$ cycle is at order $R$, and the distance between two adjacent sites is proportional to $1/R$ given $N \propto R^2$. As $R\to \infty$, $1/R\to 0$ is indeed the continuum limit.}

The value of $\tau$ itself is not as important as the scaling, which will be the crucial ingredient in what follows, since setting 
\begin{equation}
    \tilde{\tau} ~=~ \frac{2\pi}{N} \label{eq:reno}
\end{equation}
leads to an $N$-dimensional Hilbert space, or a `qudit' as discussed in section~\ref{sec:GKP}.
Of course, the scaling of modes can also be understood in `position space' \eqref{eq:rr1p}, by noting that
\begin{equation}
   \Big[\tilde{\mu}_{-}(x_{j}^{+}),\,\tilde{\mu}_{+}(x_{k}^{+})\Big]~=~i\tau\,  \delta_{jk} ~\to~ i\tau \frac{2\pi}{N}\,\delta(x-x')\,\quad\textrm{for}~ N\to\infty
   \label{eq:conc}
\end{equation}
as we introduce the continuum limit of the field, namely inserting an oscillator at each point $x^+ \in [0,2\pi)$. We will see that the symplectic discrete symmetries of \eqref{eq:rr1p} also become continuous. In the continuum limit the field acquires weight $h=1/2$ in the sense that $\tilde{\mu}(x^+)\sqrt{dx^+}$ is invariant under reparametrizations of the circle $\textrm{Diff}\,S^1$, which also leave the commutation relations \eqref{eq:conc} invariant. 

It is worth emphasizing that the same construction can be approached from a completely different `bottom-up' perspective, namely that of Celestial CFT as we will discuss in section~\ref{sec:celestialcode}.  For general spacetimes, the general form of the incidence relation \cite{Adamo:2021bej} supplements global modes $k=\pm1/2$ with what a tower of conformal descendants as shown in (\ref{eq:wqew}). We refer to them as soft hair, since it will be shown in section \ref{sec:celestialcode} that they carry supertranslation charge. Indeed, the soft hairy modes are obtained by acting with the Virasoro generators on the global ones.

\subsection{\texorpdfstring{$N$-qudit System}{N-qudit system} }\label{sec:sym}

In the previous section, we constructed a finite-dimensional code subspace where soft fluctuations can be corrected. We are now ready to extend that discussion and consider a system of $N$ independent oscillators. As explained, this corresponds to a spin chain made of $N$ qudits, which we expect to flow towards the CFT as $N\to \infty$.

Recall that each degree of freedom is labeled as $\mu_{\alpha}^{(k)}$ ($\alpha=\pm$). The system can then be written as 
\begin{equation}
    \Big[ \mu_{\a}^{(k)},\,\mu_{\b}^{(l)}\Big]  ~=~ i\,\tau\,\epsilon_{\a\b}\,\d^{k+l}\quad,\quad -\frac{N-1}{2} ~\leq~ k,l ~\leq~ \frac{N-1}{2} ~~.
    \label{equ:mu-mode-exp}
\end{equation}
The overall degrees of freedom live in a product Hilbert space $\mathcal{H}^{\otimes N}$. Further using the encoding construction, setting $\tau=2\pi/N$, we will restrict $\mathcal{H}_{\textrm{code}}$ to be $N$-dimensional as well.

Our approach to constructing a CFT and Hamiltonian is to understand the symmetries of the system and reinterpret them as currents. The system \eqref{equ:mu-mode-exp}, regarded as the physical space of harmonic oscillators, has a continuous symmetry corresponding to ${\rm Sp}(2N,\mathbb{R})$, the symmetry of the unquantized phase space. Quantization, namely encoding our qudits, then breaks ${\rm Sp}(2N,\mathbb{R})\to {\rm Sp}(2N,\mathbb{Z})$. Let us inspect the `classical' symmetry ${\rm Sp}(2N,\mathbb{R})$ first since it will have direct consequences for the CFT. 

We could very well redo the analysis of ${\rm Sp}(2N,\mathbb{R})$ following \cite{Guevara:2023tnf}, but we will take here an alternative route that will bridge the gap with the CFT construction. Indeed, we have pointed out in \cite{Guevara:2023tnf} that the symplectic group can be phrased in terms of `K\"ahler' and `non-K\"ahler' transformations. Roughly speaking, the non-K\"ahler ones are related to Virasoro symmetry, while the K\"ahler ones are related to a color $w_{1+\infty}\approx U(N\to \infty)$ symmetry. 

To introduce them, consider momentarily the field $\tilde{\mu}_{\a}(x^+)$ defined in \eqref{equ:truncated-mode-exp} along the celestial torus. We introduce two commuting generators $L_0,\bar{L}_0$ as 
\begin{equation}
    \begin{split}
        \big[L_0, \tilde{\mu}_{\a}(x^+) \big] ~&=~ -i\,\partial_{x^+}  \tilde{\mu}_{\a}(x^+) ~~,\\
     \big[\bar{L}_0, \tilde{\mu}_{\pm}(x^+) \big] ~&=~ \pm \frac{1}{2}  \tilde{\mu}_{\pm}(x^+) ~~,
    \end{split}
\end{equation}
(with this definition we can write $T=e^{2\pi i L_0}$ in \eqref{eq:tcn} as expected). According to the expansion \eqref{equ:truncated-mode-exp}, their action on the modes is then
\begin{align}
    \big[L_0, \mu_{\a}^{(k)} \big] ~&=~ k\, \mu_{\a}^{(k)} \label{eq:fss2}~~,\\
     \big[\bar{L}_0, \mu_{\pm}^{(k)}\big] ~&=~ \pm \frac{1}{2}\, \mu_{\pm}^{(k)}~~. \label{eq:fss}
\end{align}
We can now use the symplectic algebra \eqref{equ:mu-mode-exp} to derive a canonical form of the generators. One finds (setting $i\tau=1$ for simplicity here)
\begin{align}
        \bar{L}_0 ~&=~ \frac{1}{2} \sum_{k=-\frac{N-1}{2}}^{\frac{N-1}{2}} :\mu^{(k)}_{(+} \mu^{(-k)}_{-)}: \label{eq:filss} ~~,\\
    L_0 ~&=~  \sum_{k=-\frac{N-1}{2}}^{\frac{N-1}{2}} k\,:\mu^{(k)}_+ \mu^{(-k)}_-: ~~.
    \label{eq:filss2}
\end{align}
Here the symmetric bracket is defined as $A_{(\a}B_{\b)}=\frac{1}{2}[A_{\a}B_{\b}+A_{\b}B_{\a}]$ and inside the normal ordering $\mu_+$'s always sit at the left of $\mu_-$'s. 
Now, a general symplectic transformation acting linearly can always be represented via a generic matrix $T_{kj}^{\a\b}$, i.e. we have the generator
\begin{equation}
    \mathcal{T} ~=~T_{kj}^{\a\b} \,:\mu^{(k)}_{\a} \mu^{(j)}_{\b} : ~~,
    \label{equ:generalT}
\end{equation}
where the sum occurs over all the modes. We define holomorphic transformations just as in CFT:
\begin{equation}
    \left[\bar{L}_0,\,\mathcal{T}\right]~=~0~~.
\end{equation}
From the action \eqref{eq:fss} we see that any monomial $:\mu_+^{(k)}\mu_-^{(j)}:$ commutes with $\bar{L}_0$, hence holomorphic transformations are given by the ${\rm GL}(N)$ generators
\begin{equation}
    \mathcal{M} ~=~ M_{kj}\, :\mu_+^{(k)}\mu_-^{(j)}: ~~.
    \label{eq:mants}
\end{equation}
This ${\rm GL}(N)$ is precisely ${\rm Sp}(2N,\mathbb{R})\cap O(N,N)$, where $ O(N,N)$ is the Kleinian group that preserves the quadratic form \eqref{eq:filss}. Also note that $L_0$ is a holomorphic transformation with $M_{kj}=k\delta_{k+j}$.

In a similar way, we introduce antiholomorphic transformations via the relation 
\begin{equation}
    \left[L_0,\,\mathcal{T}\right]~=~0~~.
\end{equation}
From the action \eqref{eq:fss2} we see that any monomial $:\mu_{\a}^{(k)}\mu_{\b}^{(-k)}:$ commutes with $L_0$, hence these transformations have the form
\begin{equation}
    \tilde{\mathcal{{M}}} ~=~\sum_k\, M_k^{\alpha \beta}\,:\mu_{\a}^{(k)}\mu_{\b}^{(-k)}: 
    \label{eq:mtsa}
\end{equation}
with $M_{-k}^{\a\b}=M_{k}^{\b\a}$. They are simply the ${\rm GL}(2)_{\textrm{right}}$  K\"ahler transformations of the 2-qubit system, for which the $N$-qudit system has $N/2$ copies. Indeed, $L_0$ generalizes the K\"ahler operator $R^2$ there to a sum of $N/2$ such quadratic forms with different prefactors.\,\footnote{In twistor variables it is easy to understand the emergence of $N/2$ copies as $N\to \infty$. For each of the fibers $\pi^{-1}(z)\approx \mathbb{R}^2$ of the twistor bundle there is a canonical form $d\mu_\alpha \wedge d\mu^\alpha$. This will be promoted to the \textit{loop} of the symplectic algebra, $Lw_{1+\infty}$, in the next section, which contains chiral Lorentz generators.} Again, note that from this perspective $\bar{L}_0$ is an antiholomorphic transformation with $M_k^{\alpha \beta}= \frac{1}{2}\delta^{\a}_{(+}\delta^{\b}_{-)}$.

Aiming towards the CFT description, our task is now to obtain `holomorphic factorization'. Namely, we seek to enlarge the generators $L_0$ and $\bar{L}_0$ as much as possible into two commuting algebras $\mathcal{V}_{\textrm{left}}\times \mathcal{V}_{\textrm{right}} $ which will represent the chirality of the CFT. Inspired by the Iwasawa/Cartan decomposition of semisimple Lie algebras, we seek to split the generators $:\mu^{(k)}_{\a} \mu^{(j)}_{\b} :$ into commuting symmetric and antisymmetric parts.\,\footnote{By doing so, one can easily see that the generic quadratic operator (\ref{equ:generalT}) can be decomposed in the following way 
\begin{equation}
    T^{\a\b}_{kj} ~=~ T_{kj}\,\epsilon^{\a\b} ~+~ M^+_{kj}\,\d^{\a}_+\d^{\b}_+ ~+~ M^0_{kj}\,\d^{\a}_{(+}\d^{\b}_{-)} ~+~ M^-_{kj}\,\d^{\a}_-\d^{\b}_- ~~, 
\end{equation}
where $T$ is antisymmetric and $M$'s are symmetric matrices. Direct counting shows that the total number of independent components of $T^{\a\b}_{kj}$ is $N(2N+1)$, which is exactly the dimension of ${\rm Sp}(2N,\mathbb{R})$. } 
Indeed, the symmetric combination $:\mu_{(\a}^{(k)}\mu_{\b)}^{(-k)}: +:\mu_{(\a}^{(j)}\mu_{\b)}^{(-j)}: $ in \eqref{eq:mtsa}  is a ${\rm SL}(2,\mathbb{R})$ generator, which then commutes with the antisymmetric part $:\mu_{[+}^{(k)}\mu_{-]}^{(j)}:$ in \eqref{eq:mants}.  This shows that the following generators commute
\begin{align}
    w^{(2)}_{\alpha \beta}~&=~\sum_k :\mu_{(\a}^{(k)}\mu_{\b)}^{(-k)}: \,~\qquad \in \mathcal{V}_{\rm right} ~~,\label{eq:w2a}\\ 
    T~&=~\sum_{k,j} T_{kj} :\mu_+^{(k)}\mu_-^{(j)}: \,\qquad \in \mathcal{V}_{\rm left} ~~.
\end{align}
The notation will become clear in the next section. Here $T_{kj}$ is a general antisymmetric matrix, which implies that $T$ has the same action on $\mu_+^{(k)}$ and  $\mu^{(k)}_{-}$. Note that $ w^{(2)}_{+-}=
\bar{L}_0$ and that we have the global ${\rm SL}(2,\mathbb{R})$  algebra with $\bar{L}_+= w_{++}^{(2)}$ and $\bar{L}_-= w_{--}^{(2)}$.  Analogously, we can generalize $L_0$ with the particular antisymmetric combination  $T_{kj}=\delta_{k+j-m}(\frac{m}{2}- k)$. Denoting this action by $L_m$, we have 
\begin{equation}
    L_m~=~\sum_{k} \left(\frac{m}{2}-k\right) :\mu_+^{(k)}\mu_-^{(m-k)}:  ~~,
    \label{eq:lmdisc}
\end{equation}
which yields
\begin{equation}
    \left[L_m, \mu_{\a}^{(k)} \right] ~=~ -\left(\frac{m}{2}+k\right) \mu_{\a}^{(k+m)} ~~. 
    \label{eq:Lm-action}
\end{equation}
Recalling the truncation of the modes $|k|\le\frac{N-1}{2}$, we can identify this as a finite-dimensional version of the Virasoro group\,\footnote{In the $N\to\infty$ limit, the Virasoro algebra becomes clear as shown in (\ref{eq:Tz-symplecticboson}).}, containing $2N-3$ generators that act nontrivially on the phase space variables $\mu_{\a}^{(k)}$.\,\footnote{We leave a detailed discussion on this truncated version of Virasoro in appendix~\ref{appen:dvir}.}
When we take the large $N$ limit ($N\to \infty$), the truncated $\widehat{{\rm Vir}}_{\rm left}$ becomes the actual ${\rm Vir}_{\rm left}$.

To close this part of the section, let us point out that if we lift the restriction of linear transformations, we can actually write down non-linear generators that commute with $L_0$:
\begin{equation}
    w^{(p)}_{\a_1\ldots \a_p} ~=~ \sum_{\sum k_i =0}\,:\mu^{(k_1)}_{(\a_1}\cdots \mu^{(k_p)}_{\a_p)}: \qquad (p \textrm{ even})
\label{eq:wpmg}\end{equation}
extending $w^{(2)}_{\a\b}$ in \eqref{eq:w2a}. Turns out, the algebra of these generators is known as $w_{\infty}$~\cite{Guevara:2022qnm}. A discretization of this algebra will be explained in the following.

\subsubsection{Stabilizer Space and Discrete Symmetry}

Now we are ready to take the quantization into account and encode these $N$ qudits in a similar manner as section~\ref{sec:GKP}. 
As we have seen above, introducing the code breaks the symmetry group down into a discrete and finitely generated subgroup, namely the Clifford group. In what follows, we will see how this works. 

First, note that the phase space is now $2N$-dimensional, hence the finite-dimensional code subspace is generated by $2N$ stabilizers defined as follows:
\begin{equation}
S_{\pm}^{(k)} ~:= ~e^{i\,N\,{\mu}_{\pm}^{(k)}} ~= ~e^{i\,N\,s_{\pm}^{(k)}} ~~,
\label{equ:Sk}
\end{equation}
which are $N$ copies of (\ref{eq:stabilizer}). 
The error syndrome $s_{\pm}^{(k)}$ is the eigenvalue of ${\mu}_{\pm}^{(k)}$ and living in the lattice $s_{\pm}^{(k)}\sim s_{\pm}^{(k)} + \frac{2\pi}{N}$. The code subspace is defined as the simultaneous eigenspace of $S_{\pm}^{(k)}$'s with $s^{(k)}_{\pm}=0$, i.e. 
\begin{equation}
S_{\pm}^{(k)}~=~1~~.
\label{eq:sta}
\end{equation}

The logical operators that act on the code subspace are generalized Pauli gates as the $U(N)^{\otimes N}$ generators obtained from $N$ copies of (\ref{equ:gate}):
\begin{equation}\label{eq:clifgates}
     G^{(-\frac{N-1}{2})}_{\lambda_{-\frac{N-1}{2}}}\otimes\ldots \otimes G^{(\frac{N-1}{2})}_{\lambda_{\frac{N-1}{2}}} = \exp \left(i\,\sum_{k=-\frac{N-1}{2}}^{\frac{N-1}{2}} [\lambda_k \mu^{(k)}]\right)~~.
\end{equation}

Given this $N$-qudit system, the $U(N^N)$ transformations, acting by conjugation on \eqref{eq:clifgates}, that preserve its tensor product structure, form the Clifford group. 
These are nothing but the quantized versions of the holomorphic and antiholomorphic symmetries described above. The quantization/discretization ensures that the stabilizer condition \eqref{eq:sta} is valid in the new frame, via an argument analogous to \eqref{eq:2szt}. Because of this, we can alternatively characterize the Clifford group as preserving the space of stabilizer states, i.e. codes.

The generators of the Clifford group including SUM/CNOT, Fourier, and Phase gates~\cite{Gottesman:1998hu} follow from exponentiating the generators we have already discussed:

\begin{itemize}
    \item SUM/CNOT Gate (in $\mathcal{V}_{\textrm{left}}$): For some fixed value of $k$ and $l$, 
the SUM gate is given by the following  transformation in terms of the phase space variables
\begin{equation}
   \mu_+^{(k)} ~\mapsto~ \mu_+^{(k)} + \mu_+^{(-l)} ~~,~~
    \mu_-^{(l)} ~\mapsto~ \mu_-^{(l)} - \mu_-^{(-k)}  ~~.
\end{equation}
This generates the discrete symplectic transforms that only act on the mode indices, corresponding to (\ref{equ:mode-index-transf}) for the continuous case. 
\item Fourier and Phase Gates (in $\mathcal{V}_{\textrm{right}}$):   
Both the Fourier gate and phase gate are elements of ${\rm SL}(2,\mathbb{Z})_{\rm right}$ for a single qudit:
\begin{equation}
    \begin{split}
        \text{Fourier gate F: }& ~~\mu^{(k)}_+ ~\mapsto~ \mu^{(-k)}_- ~~,~~ \mu^{(-k)}_- ~\mapsto~ -\,\mu^{(k)}_+ ~~,\\
        \text{Phase gate P: }& ~~ \mu^{(k)}_+ ~\mapsto~ \mu^{(k)}_+~~,~~
    \mu^{(-k)}_- ~\mapsto~ \mu^{(-k)}_- -\mu^{(k)}_+  ~~.
    \end{split}
\end{equation}
\end{itemize}
Finally, as we saw above, ${\rm Sp}(2N,\mathbb{R})$ also contains the truncated analog of superrotations ($\widehat{\rm Vir}_{\rm left}$). Hence the symmetry breaking implies that superrotations are quantized. As $N\to\infty$, discrete superrotations are generated by
\begin{equation}
    \exp{\oint\frac{dz}{2\pi i}\, g(z)T(z)} ~=~ \exp{\sum_{k}g_{k}L_{k}} ~~,
\end{equation}
where $g_{k}\in\mathbb{Z}$. 

Besides these transformations, it is important to note that the $U(N)^N$ operators \eqref{eq:clifgates} also preserve the stabilizer condition \eqref{eq:sta}.  In fact, they are logical operations in the sense of QEC. They will be the main object we will study next section.

\section{Celestial CFT from QECC
}\label{sec:celestialcode}

In the last section, we have described an $N$-qudit system and its symmetries. Due to the relation between operator algebra and stabilizer group, symmetry analysis is crucial for quantum error correction to hold. Moreover, if one were to follow the idea that gravitational theories can be described by QECC with physical states being boundary states, the symmetries of the phase space should provide the guideline for the boundary theory. 
Indeed, we will see that as $N\to\infty$ the emergence of Virasoro and $w_{1+\infty}$ suggests the boundary theory being celestial CFT (CCFT).  The latter is a result of the isomorphism between the large $N$ limit of ${\rm SU}(N)$, i.e. the operator algebra in the code space, and the diffeomorphism group of a 2D torus, or equivalently, $w_{1+\infty}$~\cite{Hoppe:1988gk}
\begin{equation}
    \lim_{N\to \infty}\,{\rm SU}(N) ~\cong~ {\rm Diff}_{\rm T^2} ~=~ w_{1+\infty} ~~.
    \label{equ:largeNiso}
\end{equation} 
These algebras find applications in many contexts, in particular, the twistor space realization of celestial $w_{1+\infty}$~\cite{Adamo:2021bej,Adamo:2021lrv,Bu:2022iak}
and color-kinematic dualities~\cite{Cheung:2022mix,Monteiro:2022lwm,Guevara:2022qnm}. One can see how this isomorphism works from ${\rm SU}(N)$ representations in appendix~\ref{appen:SU(N)}.

Recent discoveries in asymptotically flat gravity have firmly established the equivalence between Ward Identities and soft graviton theorems of scattering process~\cite{Strominger:2017zoo}. We may ask: How many soft theorems are there and do they (over) constrain gravitational observables such as the ${\cal S}$-matrix? 

By analyzing collinear singularities in a conformal eigenbasis it was shown in \cite{Guevara:2019ypd} that there exists an infinite hierarchy of soft currents. Correlation functions inserted with these currents are interpreted as Ward identities/soft theorems in CCFT. In order for this hierarchy not to over-constrain the ${\cal S}$-matrix, an integrability condition is required. It turns out, such a condition emerges in the form of a symmetry algebra formed by the infinitely many soft currents. The old and famous BMS supertranslation and Virasoro group on the celestial sphere at null infinity (i.e. superrotations) fill the first two entries in the algebra. The infinite symmetry enhancement was identified as $w_{1+\infty}$ symmetry via 2D CFT treatment~\cite{Guevara:2021abz,Strominger:2021mtt} and later traced back to the symmetry of self-dual gravity~\cite{Adamo:2021lrv}.

In light of these results, the relation (\ref{equ:largeNiso}) is in line with the anticipation that in the large $N$ limit, celestial CFT emerges as the boundary theory from the code. In particular, the physical states correspond to graviton operators in CCFT. The aim of this section is to construct such a code. We will argue that by appropriately embedding a qudit, one can essentially  `resolve' the vacuum degeneracy arising from IR effects, as the encoding indeed protects the logical subspace from small i.e. ``soft'' errors.

Mathematically speaking, the code construction relies on the twistor space realization of the $w_{1+\infty}$ hierarchy of soft currents~\cite{Adamo:2021bej,Guevara:2022qnm}. Our procedure is as follows. 
We begin with deriving the supertranslation eigenstate ${\mg}_{\eta}$ bottom-up from the celestial CFT assumptions in section~\ref{sec:gen-supertransl}. 
In what follows, we refer to them as \textit{hard states}. 
The supertranslation charges that hard states carry are modes of the generalized wavefunction $\eta(z)$ living on the celestial torus. In section~\ref{sec:mode-exp}, we reproduce the supertranslation eigenstate top-down from the large $N$ limit of the $N$-qudit code. 
It becomes clear that the hard states are essentially living in the many-oscillator system. Based on section~\ref{sec:GKP}, we are now able to construct a stabilizer code where physical states are CCFT supertranslation states. It turns out that the logical states are the hard states with quantized soft hair as shown in section~\ref{sec:encoding}.  
Finally in section~\ref{sec:error=soft} we show that the correctable errors acting on the code subspace are soft graviton insertions.

\subsection{Bottom-up: Supertranslation Hair}\label{sec:gen-supertransl}

To introduce the equivalence, we first provide a `bottom-up' approach that leads to the formulation of QEC based on the usual assumptions of CCFT. This will also help to set the stage for the formalism. Let us recall the basic setup of Celestial CFT  (we refer to \cite{Guevara:2022qnm} for further details and relevant notation):  The aim is to compute correlation functions, involving graviton insertions as well as matter, that reproduce a gravitational ${\cal S}$-matrix. In this paper, we will be interested in the positive-helicity sector of graviton states for simplicity, while we anticipate that the formalism can be extended to more exotic CFTs.

Asymptotically flat scattering can be described in terms of graviton momentum eigenstates which we will denote as $G_{\tilde{\lambda}}(z)$. 
Their momenta read $p_{\a \dot{\a}}=\tl_{\a} \lambda_{\dot{\a}}$ with\,\footnote{We can raise indices and write $ \tl^{\a}(\bz) = \omega(\bar{z}, \, 1 )$. The spinor index $\a=-,+$ is raised/lowered by the spinor metric as follows: $A_{\a}\epsilon^{\a\b}=A^{\b}$ and $A^{\a}\epsilon_{\a\b}=A_{\b}$ with $\epsilon_{-+}=\epsilon^{+-}=+1$. 
Namely, $\l^-=\l_+$ and $\l^+=-\l_-$.}
\begin{align}
   \tl_{\a}(\bz) ~&=~ \omega\,(-1, \, \bar{z}) ~~, \label{equ:aal}\\
  \lambda_{\dot{\a}}(z) ~&=~ (1,\, z) ~~\Leftrightarrow~~ \langle \lambda + \rangle ~=~1~~,~~\langle  \lambda - \rangle ~=~ z ~~,  \label{equ:aal2}
\end{align}
where $\omega$ denotes the energy. In Kleinian signature, $\tl_{\a}$ and $\lambda_{\dot{\a}}$ are independent and reflect the factorization of the Lorentz group into two copies of ${\rm SL}(2,\mathbb{R})$.  To describe gravitons as these CFT states, we will proceed from now on with a fixed little group scaling.

The main idea to motivate the code is to accommodate for supertranslation charges, also referred to as soft hair. These exist naturally in the boundary as a consequence of the BMS large gauge transformation  \cite{Strominger:2017zoo}. In the language of CCFT, these charges are derived from the leading soft graviton as modes $P_{m\alpha}$, with $\alpha=\pm$  and $m\in \frac{1}{2}+\mathbb{Z}$. They act on a momentum eigenstate  $G_{\tl}(z)$ via
\begin{align}
\big[P_{m{\alpha}},G_{\tl}(z)\big] ~=~ i\,\tl_{{\alpha}}\,z^{m-1/2}\,G_{\tl}(z) ~~.
\label{eq:asd}
\end{align}
This means that all supertranslation charges are proportional to $\tl_{{\alpha}}$ and hence redundant. This is expected since momentum eigenstates cannot account for infinitely many quantum numbers and must lie in a particular superselection sector of the theory.

We would like to consider states with independent supertranslation charges for $m\ge 1/2$ which can encode quantum information. 
To achieve this, let us look more carefully at \eqref{eq:asd} in the form of a local OPE. Introducing supertranslation currents $P_{\pm}(z)$ as in \cite{Fotopoulos:2019vac,Himwich:2020rro}\,\footnote{The supertranslation currents are related to the leading soft graviton $H^{1}(z,\bz)$ by the following $P_{+}(z)=H^{1}(z,0)$ and $P_{-}(z)=[\bar{L}_{-1},H^{1}]_{\bz=0}$.} , they read
\begin{equation}
    \begin{split}
        P_{-}(z)\,G_{\tl}(w) ~& \sim~i\,\frac{\omega}{z-w}\,G_{\tl}(w) ~~, \\
P_{+}(z)\,G_{\tl}(w) ~&\sim~ -\,i\,\frac{\omega}{z-w}\,\bar{w}\,G_{\tl}(w)~~.
\label{eq:sda}
    \end{split}
\end{equation}
We thus see that $P_{\pm}(z)$ can be interpreted as $h=1$ (left-moving) currents~\cite{Costello:2022wso}. They can be interpreted as local generators, or global ones if we provide a meromorphic smearing function $\eta^{\pm}(z)$. To do this, it is convenient to rescale the generators into a $h=1/2$ field, 
\begin{equation}
\mu_{-}(z)~\equiv~  P_{-}(z) \langle \lambda +\rangle ~~,~~ \mu_{+}(z) ~\equiv~ P_{+}(z) \langle \lambda+\rangle ~~,
\label{eq:muisp}
\end{equation}
where we recall that $\langle  \lambda +\rangle =1 $ in the frame \eqref{equ:aal2}. Here $\langle \lambda +\rangle$ is picking a reference spinor $|r\rangle =|+\rangle$, which will turn out to be natural momentarily. With these in mind, we can introduce a meromorphic smearing function $\eta^{\pm}(z)$ of weight $h=1/2$ such that the global generator of \eqref{eq:sda} reads
\begin{equation}
    \mg_{\eta}~=~\exp\left(i\oint \frac{dz}{2\pi i} [ \eta^{+}(z) \mu_{+}(z) +\eta^{-}(z) \mu_{-}(z) ]\right) ~=~ \exp{\left(i\oint \frac{dz}{2\pi i}[\eta(z)\mu(z)]\right)} ~~.
    \label{equ:mg-eta}
\end{equation}
We describe the contour in the next subsection. 
The generalized state $\mg_{\eta}$ is a cloud of soft gravitons with possible singularities/insertions localized near the origin $\langle \lambda - \rangle = 0$. This is a particular case of coherent states of gravitons considered in e.g. \cite{Crawley:2023brz}. Our goal in the context of error correction, is to think of the exponentiation as a finite supertranslation which will be dual to a putative unitary gate. It turns out this interpretation is possible if we introduce a central term in the $P_{+}P_{-}\approx \mu_{+}\mu_{-}$ OPE so that gravitons can be paired. This term has arisen recently from many contexts~\cite{Adamo:2021bej,Adamo:2021lrv,Bu:2022iak,Guevara:2022qnm} \footnote{Most notably, the central term has been identified as the first level generator of the $w_{1+\infty}$ tower, dubbed $w^{(1)}$. }. Following the identification \eqref{eq:muisp}, it turns into the OPE
\begin{equation}
    \mu_{\a}(z_1)\,\mu_{\b}(z_2)~\sim~\frac{i\,\tau}{z_{12}}\,\epsilon_{\a\b} ~~,
    \label{eq:mumuope}
\end{equation}
where we have introduced $\tau$ as a scale parameter tracking Wick contractions. Assuming that $\eta(z)$ has localized singularities we find, combining the above formulas:
\begin{equation}
    P_{\alpha}(z)\,\mg_{\eta} ~ \sim~i\tau \,\eta_{\a}(z)\,\mg_{\eta} ~~.
\end{equation}
In the next subsection we also introduce the quantities $\eta_{\alpha}^{(m)}$ corresponding of modes of  $\eta_{\alpha}(z)$. With this definition we have
\begin{equation}
\big[P_{m{\alpha}},\,\mg_{\eta}\big]~=~ i\,\eta_{\alpha}^{(m)}\,\mg_{\eta} ~~.
\end{equation}
We see that $\mg_{\eta}$ is the desired state carrying an infinite set of soft hair, thereby generalizing the momentum states $G_{\tl}$. 
Indeed, equation (\ref{eq:sda}) follows by picking $\eta(z)=\frac{\tl}{z-w}$, which then yields
\begin{equation}
\mg_{\eta=\frac{\tl}{z-w}} ~=~ \exp{\left(i\oint \frac{dz}{2\pi i}[\eta(z)\mu(z)]\right)} ~=~ e^{i\,[\tl\mu(w)]} ~=~ G_{\tl}(w) ~~. 
\label{equ:Gtl}
\end{equation}
In the above formulas, we have adopted normal ordering implicitly. We can interpret these states as vertex operators, where $\mu(z)$ plays the role of the free field as in the standard sigma model. Indeed, the expectation is that due to the new central term \eqref{eq:mumuope} this field behaves as a gravitational Goldstone mode.\,\footnote{This is closely related to Wilson lines, see e.g. discussion in \cite{Bu:2022dis,Crawley:2023brz}.}

Performing the full Wick contraction leads to the so-called Weyl algebra: 
\begin{equation}
    \begin{split}
\mg_{\eta_{1}}\,\mg_{\eta_{2}} ~&
=~\exp\left(\oint \frac{dz_1dz_2}{2\pi i}\eta_{1}^{\alpha}\eta_{2}^{\beta}\langle\mu_{\alpha}(z_1)\mu_{\beta}(z_2)\rangle\right)\,\mg_{\eta_{1}+\eta_{2}}\\
~&=~\exp\left(i\tau\oint \frac{dz}{2\pi i}[\eta_{1}\eta_{2}]\right)\,\mg_{\eta_{1}+\eta_{2}}\\
 ~& \approx~\left(1+i\tau\oint \frac{dz}{2\pi i}[\eta_{1}\eta_{2}]\right)\,\mg_{\eta_{1}+\eta_{2}}~~.
\end{split}
\label{equ:mg-OPE}
\end{equation}
This is a new form of the CCFT algebra which, after a contraction $\tau\to 0$, can be seen to match the  $w_{1+\infty}$ algebra expected in the positive helicity sector\,\footnote{Following \cite{Guevara:2022qnm}, to take the strict limit $\tau\to0$, we need to perform a contraction
of the algebra $\mg\to\tau \mg$.}. Next, we will explicitly rediscover this algebra, and the corresponding CFT, from the perspective of stabilizer codes.

\subsection{\texorpdfstring{Top-down: CFT from $N\to \infty $}{Top-down: CFT from N to infty }}\label{sec:mode-exp}

We will now reconstruct the above results from the $N\to \infty$ limit of the stabilizer code. The main guideline to reach a CFT is simply to identify the currents associated to the symmetry algebras we described in the last section.

Let us start by considering the $N\to \infty $ limit of the oscillator field \eqref{equ:truncated-mode-exp}
\begin{equation}
  \mu_{\a}(z) ~=~ \sum_{k=-\frac{N-1}{2}}^{\frac{N-1}{2}}\frac{\mu^{(k)}_{\a}}{z^{k+\frac{1}{2}}} ~~\longrightarrow~~    \mu_{\a}(z) ~=~ \sum_{k\in \mathbb{Z}+1/2}\frac{\mu^{(k)}_{\a}}{z^{k+\frac{1}{2}}} ~~.
  \label{eq:framperp}
\end{equation}
 Here we have analytically continued the conformal coordinate $z=e^{ix^+}$, as detailed in appendix \ref{appen:modes}. We will use this continuation to define a Sigma model on $\mathbb{CP^1}$ with a bona fide OPE. Importantly, as we are inserting states on the celestial torus (a slice with $z\in\mathbb{RP}^1$) we expect physical singularities of $\mu(z)$ over the real line. This happens e.g. when two such operators are null-separated.

As explained in section~\ref{sec:sym} (see also appendix~\ref{appen:dvir}), the action of the discrete symmetry group $\textrm{Vir}_N \times U(N)$ on $\mu^{(k)}_\a$ naturally extends to the conformal fields $\mu_\a(x^+)$ and  $\mu_\a(z)$. Let us now analyze the generators in the particular frame given by \eqref{eq:framperp}. Start with the $N\to \infty$ limit of $L_m$ given in \eqref{eq:lmdisc}:
\begin{equation}
    L_m ~\longrightarrow~ \sum_{k\in \mathbb{Z}+1/2} \left(\frac{m}{2}-k\right) :\mu_+^{(k)}\mu_-^{(m-k)}: ~~,
\end{equation}
and we define 
\begin{equation}
    T(z) ~\equiv~ \sum_{m\in\mathbb{Z}}\,\frac{L_m}{z^{m+2}} ~~.\\
\end{equation}
Combining this with \eqref{eq:framperp} we can repackage $T(z)$ into
\begin{equation}
    T(z)~=~ : \mu_{[+} (z)\, \partial \, \mu_{-]} (z): ~=~ \frac{1}{2}\,\epsilon^{\a\b}\,:\mu_{\a}(z)\,\pa\,\mu_{\b}: ~~, 
    \label{eq:Tz-symplecticboson}
\end{equation}
where $\partial={\partial}/{\partial z}$. This is indeed the stress-energy tensor of a holomorphic CFT: the free symplectic boson (its OPE is the bosonic version of the familiar free fermion OPE)\,\footnote{The Virasoro algebra can be verified by computing the $T(z)T(w)$ OPE using the $\mu(z)\mu(w)$ OPE given in Eq.~\eqref{eq:mumuope}, following the standard Wick contraction procedure. Another route is to check the Jacobi identity directly using the charge action given in Eq. (\ref{eq:Lm-action}).}. The operator $T(z)$ can alternatively be derived from its discrete version $T(z_i)$ acting on the $N$-qudit system, along the lines of appendix~\ref{appen:dvir}.

Now, as it is known, the free fermion CFT has $\bar{T}=0$. However, following the previous section, we can indeed construct antiholomorphic generators from the internal $U(N)$ symmetry.  Let us start by slightly generalizing the antiholomorphic operator \eqref{eq:wpmg}:
\begin{equation}
    w^{(p)}_{\a_1\ldots \a_p,m} ~=~ \sum_{\sum k_i =m}\,:\mu^{(k_1)}_{(\a_1}\cdots \mu^{(k_p)}_{\a_p)}: ~~.
\end{equation}
Introducing
\begin{equation}
    w^{(p)}_{\a_1\ldots \a_p}(z) ~\equiv~ \sum_m\, \frac{w^{(p)}_{\a_1\ldots \a_p,m}}{z^{m+p/2}} ~~,
\end{equation}
we can repackage the generator as 
\begin{equation}
    w^{(p)}_{\a_1\ldots \a_p}(z) ~=~ :\mu_{(\a_1}(z) \cdots \mu_{\a_p)}(z): ~~.
\end{equation}
These are nothing but the generators of the loop $w_{1+\infty}$ algebra, or $Lw_{1+\infty}$. Here $z$ is the loop variable since we obtain a copy of the algebra for each mode. We refer to  \cite{Guevara:2022qnm} for the construction of the algebra OPEs in this language (see also~\cite{Adamo:2021lrv}).

Let us now discuss the OPE of the conformal field \eqref{eq:framperp}. We can immediately show it is indeed given by \eqref{eq:mumuope} assuming the field is meromorphic in the complex plane. This follows from the prescription of appendix~\ref{appen:modes}, which allows us to establish the usual equivalence between the mode commutator and the OPE, via
\begin{equation}
\begin{split}
    \Big[ \mu_{\a}^{(k)},\mu_{\b}^{(l)}\Big] ~=&~ \oint_{\mathcal{C}} \frac{dz_2}{2\pi i} \,z_2^{l-\frac{1}{2}}\,\oint_{z_2} \frac{dz_1}{2\pi i} \,z_1^{k-\frac{1}{2}}\,\m_{\a}(z_1)\,\m_{\b}(z_2)
    ~=~ i\,\tau\,\epsilon_{\a\b}\,\d^{k+l} ~~,
\end{split}   
\label{equ:mu-mode-alg}
\end{equation}
where generically the mode indices $k$ and $l$ run from $-\infty$ to $+\infty$. 
As expected, (\ref{equ:mu-mode-alg}) coincides with (\ref{equ:mu-mode-exp}). 

\begin{figure}[htp!]
    \centering
    \includegraphics[width=0.45\linewidth]{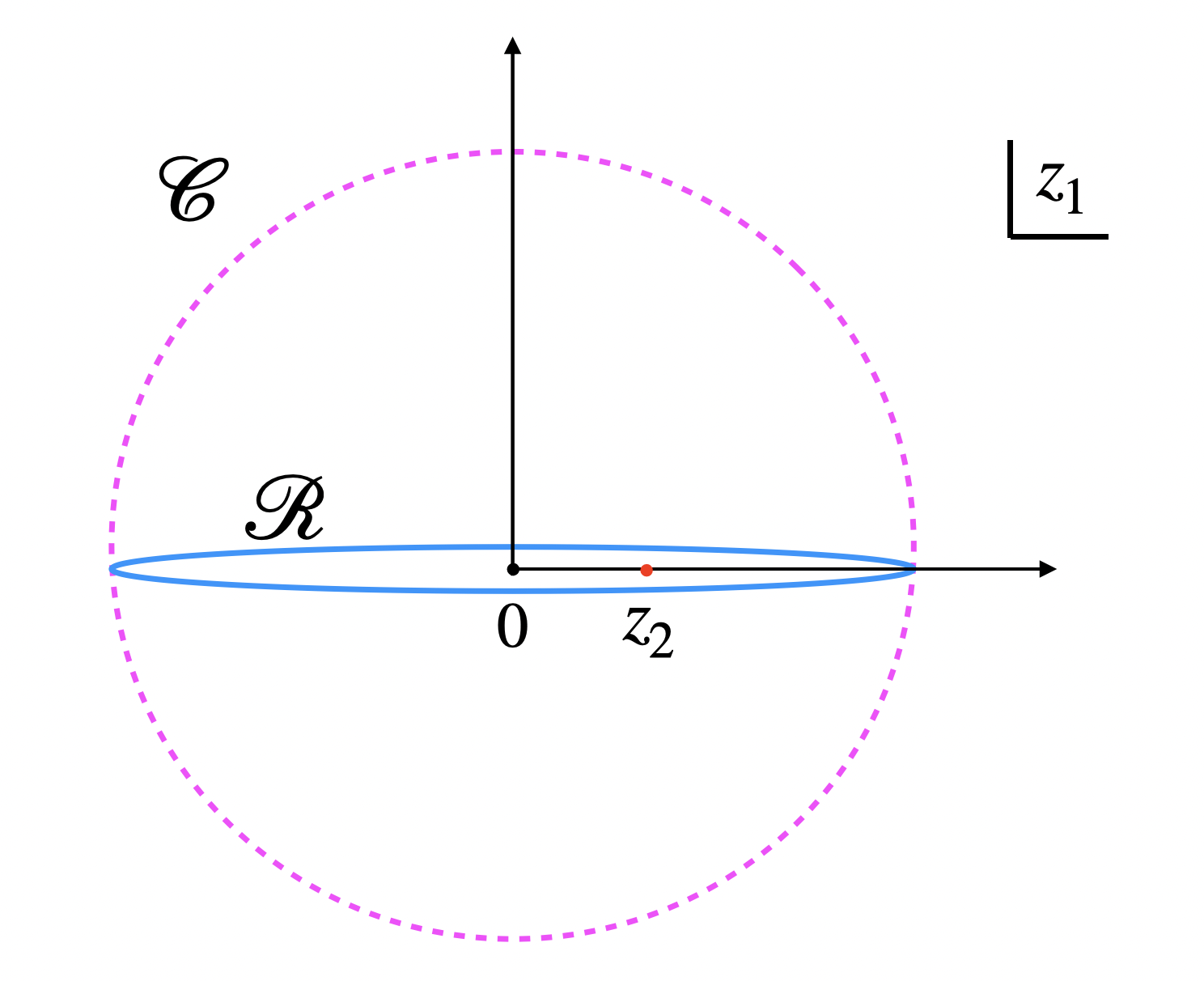}
    \caption{The choice of the contour ${\cal C}$ depicted in the complex $z_1$ plane. We choose a physical interval ${\cal R}\subset \mathbb{R}$ shown in blue and analytic continue it to the complex plane (the circular contour $\mathcal{C}$ shown in magenta). }
    \label{fig:contour}
\end{figure}

Let us give an alternative take on the choice of contour $\mathcal{C}$. We can take $\mathcal{C}$ to be the standard circular contour containing the origin or neighboring insertions, as detailed in Appendix \ref{appen:modes}. However, recall that as long as we insert operators in the celestial torus (choose, say, a patch with null coordinates $z_1,z_2 \in \mathbb{RP}^1$) we expect the OPE $\mu(z_1)\mu(z_2)$ to be singular at null separation. Thus the contours in \eqref{equ:mu-mode-alg} can be taken to hug an interval ${\cal R}\subset \mathbb{RP}^1$ as long as it contains the singularity $z_{12}=0$. We will denote this physical contour by ${\cal R}$; we return to it in section \ref{sec:encoding}. Of course, ${\cal R}$ can be analytically pushed to the complex plane, leading back to the circular contour $\mathcal{C}$ in a complexified celestial torus. We depict this in the complex $z_1$ plane in Fig.~\ref{fig:contour}.

The direct connection with the incidence relation, \eqref{eq:framperp} as well as the existence of $Lw_{1+\infty}$ fields, are characteristic of the twistor sigma model proposed in \cite{Adamo:2021bej}. As we explain further in appendix \ref{appen:modes} even though the incidence relation carries weight $h=-1/2$ the model is described by a $h=1/2$ field that can be integrated against $d^2z$:
\begin{equation}
   S ~=~ \frac{1}{\tau}\, \int_D d^2 z\, \epsilon^{\a \b}\, \mu_{\a} \bar{\partial} \mu_{\b} ~+~ S_{\textrm{source}}(\mu_{\a}) ~~,
   \label{equ:free-fermion-action}
\end{equation}
where $D\subset\mathbb{CP}^1$  can be bounded by a real slice $\partial D=\mathbb{RP}^1$ that we can identify with a null direction of the celestial torus \cite{Mason:2022hly}. Of course, the free term leads to the holomorphic condition of $\mu(z)$, as well as the OPE \eqref{eq:mumuope}.\,\footnote{This is easily checked since $\bar\partial z^{-1}= 2\pi\delta^2(z)$ in $D$, hence the propagator takes the form $\frac{1}{z_{12}}$.} The model admits different localized (delta function) sources leading to a variety of self-dual spacetimes \cite{Adamo:2021bej,Bittleston:2022nfr} that we will not discuss here. It would be interesting to understand them from the perspective of gravitational QEC, see section~\ref{sec:conclusion}.

\subsubsection*{Mode expansion}

Let us go back to the generalized states with soft hair, constructed from the formula \eqref{equ:mg-eta}.  As stated this requires $\eta^{\pm}(z)$ to have the weight $h=1/2$ and hence admits the expansion 
\begin{equation}
\eta_{\a}(z) ~=~ \sum_{k\in \mathbb{Z}+1/2}\frac{\eta^{(k)}_{\a}}{z^{k+\frac{1}{2}}} ~~,
\label{equ:mode-exp}
\end{equation}
where the real variable $z$ covers insertions at $z=0$ and $z=\infty$.  By choosing the (complexified, compact) $\mathcal{C=}S^1$ contour that only contains one of them we can expand our states in terms of their Fourier modes, which represent soft charges. We thus have
\begin{equation}
    \mg_{\eta} ~=~ \exp\left( i\,\sum_k\,\left[\eta^{(k)}\mu^{(-k)}\right] \right) ~~. \label{eq:dsgm}
\end{equation}
This is the $N\to\infty$ limit of (\ref{eq:clifgates}). This means that we can now recognize $ (\ref{eq:dsgm})$ (or (\ref{equ:mg-eta}), namely supertranslation eigenstates in CCFT) as generalized Pauli operators. This mathematical equivalence strongly suggests that the graviton modes should act as gates encoding quantum information. In the remaining sections, we attempt to flesh out this perspective. 

Final remarks are in order. By applying the Baker–Campbell–Hausdorff formula to (\ref{eq:dsgm}), we find the generalized Pauli algebra
\begin{equation}
    \mg_{\eta_1}\,\mg_{\eta_2} ~=~ \exp{\left(i\tau\,\sum_{n}\,\left[\eta_1^{(n)}\eta_2^{(-n)}\right]\right)}\,\mg_{\eta_2}\,\mg_{\eta_1}~~,\label{eq:gn1gn2}
\end{equation}
which is equivalent to the Weyl algebra \eqref{equ:mg-OPE}.
Importantly, we have performed the mode expansion around $z=0$. By enforcing supertranslation quantization\,\footnote{By supertranslation quantization we mean that the supertranslation charges that these eigenstates carry are quantized.} in section~\ref{sec:encoding}, we will break the full Virasoro symmetry down to its discrete subgroup tied to the
choice $z=0$. We anticipate that $\langle\lambda -\rangle=0$ is a privileged location in the celestial torus where fluctuations/errors can be corrected.

\subsection{Code States = Hard States}\label{sec:encoding}

Given that the supertranslation states live in a system of infinitely many oscillators, we are able to construct a quantum error-correcting code following section~\ref{sec:sym}. We will take as the physical states our generalized supertranslation eigenstates, which extend the notion of positive helicity gravitons. Indeed, \eqref{equ:Gtl} realizes the intuition that the usual momentum eigenstate is encoded in them. Furthermore, as a consequence of the OPE, we expect that the generalized operators can indeed accommodate more than one momentum eigenstate: The wavefunction $\eta(z)$ can account for the soft hair inserted by different jets of soft radiation piercing null infinity.

Let us now formulate the code precisely. As mentioned above, the key is that the supertranslation eigenstates $\mg_{\eta}$ given by \eqref{eq:dsgm} match the $N\to\infty$ limit of the phase space $U(N)^N$ operators, namely
\begin{equation}
       G^{(-\frac{N-1}{2})}_{\lambda_{-\frac{N-1}{2}}}\otimes\ldots \otimes G^{(\frac{N-1}{2})}_{\lambda_{\frac{N-1}{2}}} = \exp \left(i\,\sum_{k=-\frac{N-1}{2}}^{\frac{N-1}{2}} [\lambda_k \mu^{(k)}]\right)~~\longrightarrow~~  \mg_{\eta} ~=~ \exp\left( i\,\sum_k\,[\eta^{(k)}\mu^{(-k)}] \right) ~~.
       \label{eq:gnmat}
\end{equation}
The stabilizers take the form as (\ref{equ:Sk}), namely
\begin{equation}
S_{\pm}^{(k)} ~:=  \exp\left(N\int_{\mathcal{R}} dz\,z^{k-1/2}\mu_{\pm}(z)\right) = ~e^{i\,N\,{\mu}_{\pm}^{(k)}} ~= ~e^{i\,N\,s_{\pm}^{(k)}} ~~.\label{eq:golds}
\end{equation}
Here we are using the prescription for the contour $\mathcal{R\subset \mathbb{RP}^1}$ outlined in section \ref{sec:mode-exp}.

The code essentially follows the same procedure as in section~\ref{sec:sym}. We now aim to specify its physical meaning in the context of celestial holography. From this point of view, the code subspace of hard states is a set of states with quantized soft hair around $z=0$. These logical states can be constructed as follows. 
First, we need an appropriate stabilizer vacuum which satisfies
\begin{equation}
    S_{\pm}^{(k)}\,\mathcal{V}_{(0,0)} ~=~ \mathcal{V}_{(0,0)} ~~.\label{eq:propsf}
\end{equation}
As a generalization of (\ref{eq:quots}), we can take\,\footnote{Note that it is not the GKP vacuum since $\mathcal{V}_{(0,0)}$ is an eigenstate for both $e^{{\mu}^{(k)}_+}$ and $e^{{\mu}^{(k)}_-}$. The GKP vacuum can be written as a linear combination of it, see e.g. (\ref{equ:GKPvacuum}).
}
\begin{equation}
\mathcal{V}_{(0,0)} ~=~ \prod_{k}\,\sum_{p,q\in\mathbb{Z}}\,e^{i\,N\,p\,{\mu}_{+}^{(k)}}\,e^{i\,N\,q\,{\mu}_{-}^{(k)}} ~~.
\label{equ:vacuum}
\end{equation}
Then, using the operator-state correspondence, it is easy to construct the logical states by acting the logical operators on the vacuum
\begin{equation}
    \tilde{\mg}_{{\eta}}~=~ \mg_{{\eta}}\,\mathcal{V}_{(0,0)} ~~,\label{eq:defst}
\end{equation}
where the modes of ${\eta}^{(k)}_{\alpha}$ in \eqref{eq:gnmat} are now integers and living in the stabilizer lattice
\begin{equation}
\boxed{{\eta}_{\alpha}^{(k)} ~=~ \oint \frac{dz}{2\pi i}\,z^{k-1/2}\,{\eta}_{\alpha}(z) ~~\in\mathbb{Z}_N \times \mathbb{Z}_N  ~~.}
\end{equation}
In other words, the supertranslation charges are quantized around $z=0$. The vacuum \eqref{equ:vacuum} is constructed from these charges and thus naturally defines $z=0$ as the origin. We can think of the generalized state \eqref{eq:defst} as placing the infinite tower of $w_{1+\infty}$ charges at this insertion.\footnote{We will further assume $\eta^{1/2}_-=0$ which places the generalized state at $\bar{z}=0$. See next section.} Furthermore, note that the quantization condition is preserved as long as we translate $z=0 \to z=n\in \mathbb{Z}$. For instance, we immediately get
\begin{equation}
    \eta^{(5/2)}_{\alpha} ~\to~ \eta^{(5/2)}_{\alpha}-2n\,\eta^{(3/2)}_{\alpha}+n^2\eta^{(1/2)}_{\alpha} ~\in~\mathbb{Z}_N ~~,
\end{equation}
etc. This breaks the ${\rm SL}(2,\mathbb{R})$ action down to its discrete subgroup.

It is straightforward to check that these logical states are robust, i.e. they commute with the stabilizers, as follows\,\footnote{As operators living on the celestial torus, this means there are no contractions/OPE between them.}
\begin{equation}
    S_{\pm}^{(k)}\,\mg_{{\eta}} ~=~ e^{2\pi i\,{\eta}^{(k)}_{\pm}}\,\mg_{{\eta}}\,S_{\pm}^{(k)} ~=~ \mg_{{\eta}}\,S_{\pm}^{(k)}  ~~,
\end{equation}
and from \eqref{eq:propsf}-\eqref{eq:defst}
\begin{equation}
    S_{\pm}^{(k)}\tilde{\mg}_{{\eta}} ~=~ \tilde{\mg}_{{\eta}} ~~.
\end{equation}
It is worth emphasizing that the code subspace is essentially the Hilbert space of $N\to\infty $ qudits. 
This suggests, based on the discussion in the previous section (and \cite{Guevara:2023tnf}), that logical states can be naturally interpreted as bulk states in twistor space, thanks to the embedding of the $N$-qudit spin chain.

\subsection{Errors as Soft State Insertions}\label{sec:error=soft}

Above we have constructed the code subspace and specified that the logical states are hard states with quantized soft hair. 
What errors can be corrected in this code subspace? What is their holographic interpretation?

Recall that in this setup, a generic possible error acting on the code subspace is written as
\begin{equation}
    E_{\kappa} ~= \exp{\left(i\oint \frac{dz}{2\pi i}[\kappa(z)\mu(z)]\right)} =~ \exp\left(i\,\sum_{j}\left[\kappa^{(j)}\mu^{(-j)}\right]\right) ~~,
\end{equation}
which satisfies, using e.g. \eqref{eq:gn1gn2},
\begin{equation}
    S_{\pm}^{(j)}\,E_{\kappa} ~=~ e^{2\pi i\kappa^{(j)}_{\pm}}\,E_{\kappa}\,S_{\pm}^{(j)} ~~.\label{eq:xerr1}
 \end{equation}
The wavefunction $\kappa(z)$ is understood here as error fluctuations associated to soft radiation, as we discuss momentarily. As mentioned in section~\ref{sec:GKP}, the eigenvalues of the stabilizer $(s^{(j)}_+,s^{(j)}_-)$ are the error syndrome. In the CFT picture, they play the role of Goldstone modes.\,\footnote{Let us emphasize that in the gravitational phase space framework, the Goldstone mode serves as the symplectic partner of the soft graviton. Thus, the shift in gravitational mode can be described by the notion of Goldstone modes. }  Explicitly, the action of an error on the vacuum operator is, from \eqref{eq:propsf} and \eqref{eq:xerr1},
\begin{equation}
    E_k \mathcal{V}_{(s_+,s_-)} = \mathcal{V}_{(\tilde{s}_+,\tilde{s}_-)}
\end{equation}
This error shifts the Goldstone mode defined in \eqref{eq:golds} as follows
\begin{equation}
\left(\tilde{s}^{(j)}_{+},\tilde{s}^{(j)}_{-}\right)~=~\left(s^{(j)}_{+},s^{(j)}_{-}\right)~+~\frac{2\pi}{N}\,\left(\kappa^{(j)}_+,\kappa^{(j)}_-\right)~~.
\end{equation}
Errors are a particular case of generalized states \eqref{eq:muisp} and satisfy the $w_{1+\infty}$ algebra in the form of \eqref{eq:weylal}. This justifies their interpretation as soft gravitons, which will be elaborated in more detail shortly. 

A correctable error syndrome requires 
\begin{equation}
|\kappa^{(j)\pm}|~=~\left|\oint \frac{dz}{2\pi i}\, z^{j-1/2}\kappa^{\pm}(z)\right| ~<~ \frac{1}{2}~~.
\label{eq:sdfq}
\end{equation}
Namely, the action of states with supertranslation charges bounded as above can be corrected by adjusting $(s^{(j)}_{+},s^{(j)}_{-})$ back to zero, following a standard correction protocol. 
Recall that due to (\ref{equ:mu-mode-alg}), the unit area of the stabilizer lattice for each mode is $\tau=2\pi/N$, hence the physical fluctuations that can be corrected satisfy the following condition
\begin{equation}\boxed{~~
    |\Delta \kappa^{(j)\pm}| ~<~ \frac{1}{2}\,\sqrt{\frac{2\pi}{N}} ~~.~}
    \label{equ:QEC}
\end{equation}
Again we choose the ``symmetric" code where $\mu_{-}$ and $\mu_{+}$ can be corrected equally well for the same reason explained in footnote~\ref{ft:ex-sym-alpha}.

We refer to (\ref{equ:QEC}) as the \textit{QEC condition}. 
As $N\to\infty$, we can only correct small i.e. `infrared' fluctuations. 
For large but finite $N$ however, there is a finite energy threshold where we can encode a state. We refer to states below this threshold as (generalized) soft states. 
So far the error model follows the discussion in section~\ref{sec:qudit-error}. Recall that in the single-qudit toy model, there are only two parameters $\hat{\epsilon}_{\pm}$ describing the shifting errors. 
While in our holographic code, $\kappa^{(j)}_{\pm}$ is the mode of the wavefunction $\kappa_{\pm}(z)$. Hence more tunable parameters are involved and we will unpack the QEC condition and discuss the physical meanings of these tunable parameters below.

\subsubsection*{QEC condition}
To be more precise, consider errors being the momentum eigenstates and their supertranslation charges can be parametrized as follows
\begin{equation}
   \kappa_{\pm}(z)~=~\frac{\tl_{\pm}}{z-w}~~\Rightarrow~~ \kappa_{\pm}^{(j)} ~=~ \oint_{{\cal C}} \frac{dz}{2\pi i}\,z^{j-1/2}\frac{\tl_{\pm}}{z-w}~=~ w^{j-1/2}\tl_{\pm}~~\left(j\ge\frac{1}{2}\right)   ~~,\label{eq:cnd}
\end{equation}
where $\tl_{\pm}$ is the two-spinor defined in (\ref{equ:aal}) and the contour ${\cal C}$ is chosen to encircle both the origin and $w$ as shown in Fig.~\ref{fig:contour}. These errors are standard gravitons satisfying the loop $w_{1+\infty}$ algebra in momentum basis:
\begin{equation}
    E_{\kappa_1}E_{\kappa_2} ~\sim~ \tau\, \frac{[\tilde{\lambda}_1 \tilde{\lambda}_2]}{z_{12}} E_{\kappa_1+\kappa_2} ~~.
\end{equation}
Recalling $\tilde{\lambda}=\omega(1\,,\bar{w})$, the error correction condition \eqref{equ:QEC} has two components. For the modes \eqref{eq:cnd}, the first component reads 
\begin{equation}
    |w|^{j-1/2}\,\o\,\sqrt{N}~<~ \sqrt{\frac{\pi}{2}}~~.
    \label{equ:qec-condition}
\end{equation}
Since this holds for all $j\ge 1/2$ it requires $|w|< 1$ and
\begin{equation}
        \o~\le~ \L ~=~ \sqrt{\frac{\pi}{2N}} ~~.
    \end{equation}
Thus the energy of the inserted graviton state is bounded by a cutoff $\L$, in accordance with the standard notion of soft gravitons.

\begin{figure}[htp!]
    \centering
    \includegraphics[width=0.5\linewidth]{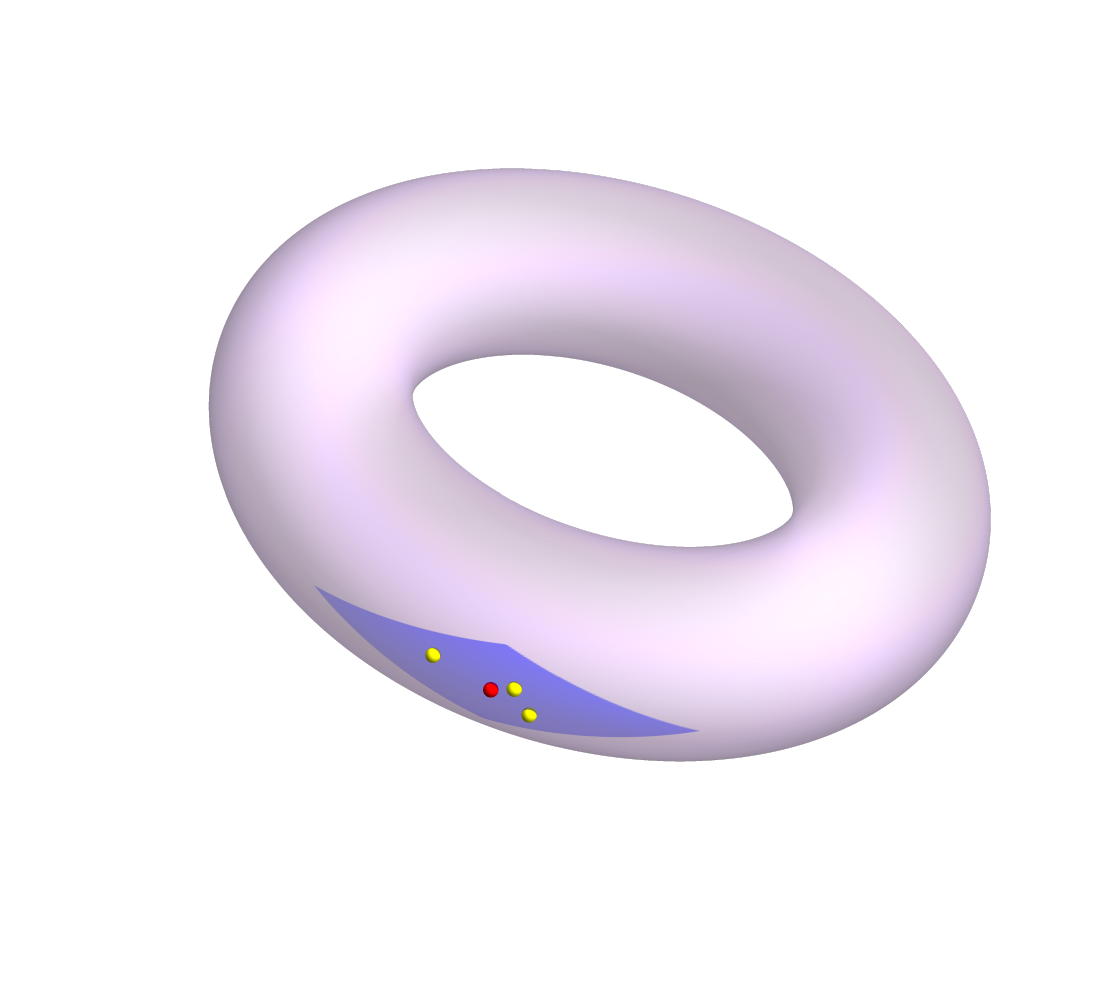}
    \caption{A diamond of the celestial torus. We assume the observer can perform measurements at around a point $z=\bar{z}=0$ (red). These shall encapsulate fluctuations induced by nearby operators $E_k$ depicted as yellow insertions.}
    \label{fig:diamond_torus}
\end{figure}

The second component of the error-correcting condition is fulfilled if $|\bar{w}|<1$, which is consistent with the CP symmetry $\bar{w} \leftrightarrow w$. We thus find
\begin{equation} \boxed{~~
   |w|\,,|\bw|~<~ 1 ~~,~~ 
    \o ~\le ~\Lambda ~~.}
    \label{equ:QEC-phys}
\end{equation}
This corresponds to a cell in the phase space of soft radiation (of volume $8\L$) where the error correction operates. More specifically, recall $(w,\bw)$ denotes the position of the soft insertion on the celestial torus. The bound corresponds to a window of the celestial torus as shown in Fig.~\ref{fig:diamond_torus}, the analog of a solid angle of the celestial sphere, where insertions of soft gravitons modify the hard state but can be corrected. Recall that the hard state is defined by charges quantized around $w=0$, which can be thought of as the position of an observer.

As $N\to\infty$ the logical subspace has a countable infinite dimensional representation corresponding to the soft charges. The correction window is a narrow IR cell $\Lambda \ll 1$ which encapsulates collinear and soft divergences in correlation functions. More precisely, we can consider dressed correlation functions of the kind
\begin{equation}
     \mathcal{A} ~=~ \langle E_{\tilde{\lambda}_1}\ldots E_{\tilde{\lambda}_m} \tilde{\mg}_{{\eta}}  \mathcal{O}_{1} \ldots   \mathcal{O}_{n} \rangle 
\end{equation}
involving an insertion of one generalized state smeared near $z=\bar{z}=0$ as well as other hard states $\mathcal{O}_i$ at distant locations. Soft gravitons with momentum inside the window \eqref{equ:QEC-phys} will contribute to collinear and IR divergences. They can be associated with error fluctuations as follows: We assume the observer is able to perform measurements at a distance $r$ from $|z|=0$. Let $\mathcal{R}$ hug the interval $[-r,r]$. We can read off the infinite tower of charges 
\begin{equation}
    \eta^{(k)}_{\alpha}~=~\int_{\mathcal{R}} \frac{dz}{2\pi i} z^{k-1/2}\eta_{\alpha}(z)\label{eq:mes}
\end{equation} and reconstruct the wavefunction $\eta(z)$ in $\mathcal{R}$. The OPE (\ref{equ:mg-OPE}) implies that the soft dressing shifts the generalized wave function of the hard state,
\begin{equation}
  \tilde{\mg}_{{\eta}} ~\to~  E_{{\tl}}\tilde{\mg}_{{\eta}} ~~,~~ {\eta}_{\pm}(z) ~\mapsto~ {\eta}_{\pm}(z) ~+~ \frac{\tl_{\pm}}{z-w} ~~
\end{equation}
(up to a phase in $\tilde{\mg}_{{\eta}}$). Hence the position $w$ of the soft insertion is encoded as the pole of the shifted wave function. Assuming $|w|\le r$ we see that equation  \eqref{eq:mes} shift as
\begin{equation}
      \eta^{(k)}_{\alpha}~\to~  \eta^{(k)}_{\alpha}+\kappa^{(k)}_{\alpha}~~,
\end{equation}
where $\kappa^{(k)}_{\alpha}$ is given in \eqref{eq:cnd}. This is detected by measuring the stabilizer
\begin{equation}
    S^{(k)}_{\pm}~=~\exp\left(N\int_{\mathcal{R}} dz\,z^{k-1/2}\mu_{\pm}(z)\right) ~~.\label{eq:somsa}
\end{equation}
For the full correlation function, we get
\begin{equation}
    \mathcal{A}~\to~ \langle\tilde{\mg}_{{\eta}'}  \mathcal{O}_{1} \ldots   \mathcal{O}_{n} \rangle\quad,\quad
    \eta'(z)~=~\eta(z)+\sum_{i=1}^m\frac{\tilde{\lambda}_i}{z-z_i} ~~.
    \label{eq:dzxaf}
\end{equation}
The energy and the positions $(\tilde{\lambda}_i,z_i)$ can be obtained by again measuring the stabilizer \eqref{eq:somsa}, gradually adjusting the radius $r$ of the interval $\mathcal{R}$, and correcting one error at a time. It is interesting to contrast this procedure with the standard local measurement, given by the insertion of a soft graviton $H(w,\bar{w})$,
\begin{equation}
    \langle H(w,\bar{w}) \tilde{\mg}_{{\eta}'}  \mathcal{O}_{1} \ldots   \mathcal{O}_{n} \rangle ~=~ \left([\eta(w) \tilde{\lambda}_{\bar{w}}]+\sum_{i=1}^m\frac{[\tilde{\lambda}_i\tilde{\lambda}_{\bar{w}}]}{w-z_i } + \ldots\right) \langle \tilde{\mg}_{{\eta}'}  \mathcal{O}_{1} \ldots   \mathcal{O}_{n} \rangle ~~,
\end{equation}
where $\ldots$ denote the contributions from $\mathcal{O}_i$. As expected this is indeed Weinberg's soft theorem, slightly extended to account for the wavefunction $\eta(z)$ of the generalized state. We conclude both measurements do indeed output the information of soft radiation, but \eqref{eq:dzxaf} stores it in an eigenvalue of a stabilizer as we anticipated.

\section{Closing Remarks}\label{sec:conclusion}

In this work, we constructed a model of celestial CFT through a quantum error-correcting code. In this code, CFT states constitute the physical Hilbert space. The logical operators that act on a subset of physical states are represented by the generalized Pauli strings for $N\to \infty$ qudits. 

Let us come back to the holographic code idea advocated in Figure \ref{fig:torus}. We have allowed the central term $\tau$ to flow according to not only the radial distance $R^2$ as $\tau\propto 1/R^2$ but also the number $N$ of qudits allocated in the code as $\tau\propto 1/N$ (equation (\ref{equ:tau-renormN})). Therefore, the QEC condition (\ref{equ:QEC-phys}) implies that as closer to the boundary, it's harder to protect the bulk states, as expected from holographic codes such as the HaPPY code~\cite{Pastawski:2015qua}. The CFT is then defined by taking the $N\to\infty$ continuum limit and manifests key features of celestial CFT. Besides the celestial chiral symmetry algebra, one is able to reproduce the graviton-graviton OPEs and tree-level MHV correlation functions in the same way as~\cite{Guevara:2022qnm} by identifying the code states as the vertex operator corresponding to the graviton states~\cite{Adamo:2021bej,Adamo:2021lrv}.  
The code itself also has a clear physical consequence: the logical states are hard states with quantized soft hair while the correctable errors are nearby soft insertions.

We close with several natural routes for future investigations. 

\paragraph{Twistor Space and Quantum Information} 
In Penrose's twistor theory, non-perturbative deformations of the symplectic form map inequivalent self-dual spacetimes. Some of these correspond to instantons that can be interpreted as self-dual deformations of black holes. From a modern perspective, these backgrounds can be introduced by adding suitable interaction terms in the sigma model \eqref{equ:free-fermion-action}. 
In QEC these maps are famous \textit{non-Clifford gates}, namely unitary transformations that do not preserve the symplectic qudit structure \eqref{eq:rr1p}-\eqref{eq:conc}. In the context of universal quantum computation, the complexity of non-Clifford gates is measured by so-called Magic \cite{Knill:2004vlm,Bravyi:2004isx,Liu:2020yso,White:2020zoz}. One pressing future direction is thus understanding the connection between deformations of twistor space and Magic. 
Celestial duality in curved backgrounds is usually carried out via reductions in twistor space. For instance, Burns space~\cite{Costello:2023hmi}, is bulk spacetime in the top-down celestial holographic duality~\cite{Costello:2022jpg}. It is expected that we can implement the ideas presented in this work for more general sigma models with a twistor target.

\paragraph{IR Divergences} 
The IR divergence in the standard\,\footnote{Contrast to the inclusive cross sections in collider physics.} massless scattering amplitudes is a long-standing issue. As mentioned in the introduction, the memory effect (equivalently the vacuum degeneracy) accounts for IR divergences. 
In order to obtain IR finite amplitudes, for example in gravitational scattering, one has to accommodate a very large set of supertranslation charges. The construction of supertranslation eigenstates has been discussed extensively~\cite{Choi:2017bna,Choi:2017ylo,Prabhu:2022zcr,Prabhu:2024zwl,Prabhu:2024lmg}. 
In section~\ref{sec:gen-supertransl} we introduce the unitary operator $\mg_{\eta}$ that carries an infinite set of supertranslation charges. 
One natural direction is to extract the infrared divergences in gravitational ${\cal S}$-matrix from $\mg_{\eta}$ and we leave it in future work. 
One potential starting point could be the following. 
Since the generalized wavefunction $\eta_{\a}(z)$ has chiral weight $h=1/2$, it generically contains $1/z$ and $O(1)$ terms. We anticipate the constant term is responsible for IR divergences based on the observation that its OPE matches with the Goldstone OPEs~\cite{Himwich:2020rro}.\,\footnote{Consider $\eta_{\a}(z)$ taking the form of $\eta_{\alpha}(z) = \tl_{\alpha}\theta(z_1-z)$, where $\theta(x)$ is the Heaviside step function. The OPE of the exponent of $\mg_{\eta}(z_1)$ matches with the Goldstone-Goldstone OPE~\cite{Himwich:2020rro} up to some constants of integration.}  
Further study is needed to fully understand its relation to the Goldstone mode and soft dressing proposed in~\cite{Himwich:2020rro}.

\paragraph{Spin Chain Models for Celestial CFT} 
So far we have been focusing on the quantum degrees of freedom. As shown in section~\ref{sec:celestialcode}, their defining algebraic structure gives the $\mu_{\a}\mu_{\b}$ OPE (\ref{eq:mumuope}) by identifying the canonical bracket with the radial quantization bracket. 
One next natural step is to take the dynamics into account and aim for concrete statements about spin chain models for celestial CFT. 
Qudits are naturally realized by models such as the clock, XY, Potts model, etc...It is anticipated that spin chain models can provide more insights into the properties of CCFT, for example, the entanglement pattern~\cite{Chen:2023tvj,Chen:2024kuq}.

\paragraph{Renormalization Scale}
Another pressing task is to understand the role of $N$ as a renormalization scale.  
Recall that in the MERA tensor network, the number of spins $N$ plays the role of a UV cutoff, and thus the additional direction corresponds to the RG flow. Further investigation is required to determine if similar statements apply in our context. 
Moreover, it would be interesting to view the finite $N$ version of the code from the quantum gravity perspective. Along this line, the discussion on the discrete version of diffeomorphism presented in appendix~\ref{appen:discrete-diff} would be a good starting point.

\section*{Acknowledgements}

It is a pleasure to thank Kevin Costello, Laurent Freidel, Daniel Harlow, Zi-Wen Liu, Sabrina Pasterski, Atul Sharma, Andrew Strominger, Diandian Wang, Zixia Wei, Carolyn Zhang and Liujun Zou for valuable discussions.
AG is supported by the Black Hole Initiative and the Society of Fellows at Harvard University, as well as the Department of Energy under grant DE-SC0007870. 
The research of YH is partially supported by the Celestial Holography Initiative at the Perimeter Institute for Theoretical Physics and the Simons Collaboration on Celestial Holography. Research at the Perimeter Institute is supported by the Government of Canada through the Department of Innovation, Science and Industry Canada and by the Province of Ontario through the Ministry of Colleges and Universities. YH's research at the University of Waterloo is supported by an Alliance Quantum Program grant funded by NSERC.

\appendix

\section{On the Mode Decomposition}\label{appen:modes}

In this appendix, we will establish a relation between the following fields
\begin{equation}
    \tilde\mu_{\a}(x^+) ~=~ \sum_k \mu^{(k)}_{\a} e^{ikx^+}~~,~~ \mu_{\a}(z) ~=~ \sum_k \frac{\mu^{(k)}_{\a}}{z^{k+1/2}}~~,~~ z=\tan{\frac{x^+}{2}}~~,
\end{equation}
where the index $k$ ranges over some finite or infinite-discrete set. For concreteness, we can focus on
\begin{equation}
    k ~=~ -\frac{N-1}{2}~,\ldots ,~\frac{N-1}{2}~~.\label{eq:eran}
\end{equation}
To obtain the relation, the idea is simply to equate the expression for the modes in both cases, namely:
\begin{equation}
  \mu^{(k)}_{\a}~=~\frac{1}{2\pi}  \int_{0}^{2\pi} dx^{+} e^{-ikx^+}\tilde\mu_{\a}(x^+)~=~\frac{1}{2\pi i} \oint_{\mathbb{RP}^1} dz z^{k-1/2} \mu_{\a}(z)~~,
  \label{eq:samemod}
\end{equation}
and argue they can be obtained from each other via contour deformation from $\mathbb{RP}^1$ to the circle $S^1$. This requires a bit of care in analyzing the $\mathbb{RP}^1$ contour. A convenient way is the homogeneous formalism of ${\rm SL}(2,\mathbb{C})$ representations.

Homogeneous coordinates $z=\tan x/2$ and $\tilde{z}=e^{ix}$ are obtained as $z=\lambda_{-}/\lambda_{+}$ and $\tilde{z}=\tilde{\lambda}_{-}/\tilde{\lambda}_{+}$, with local coordinates
\begin{equation}
    \begin{split}
        \lambda ~& =~(-1,z)~~,\\
\tilde{\lambda} ~& =~(-e^{-ix/2},e^{ix/2})~~.
    \end{split}\label{eq:2params}
\end{equation}

As explained, these are related via a ${\rm SL}(2,\mathbb{C})$ transformation.
From this frame, the functions $\mu(\lambda):=\mu(z),\,\tilde{\mu}(\tilde{\lambda}):=\tilde{\mu}(x^+)$ are promoted
to $h=1/2$ homogeneous functions via $\mu(t\lambda):=t^{-1}\mu(\lambda)$.
The advantage of this formalism is that a ${\rm SL}(2,\mathbb{R})$ transformation
$\Lambda$ is implemented via $\mu'(\lambda):=\mu(\Lambda\lambda)$.
Introduce
\begin{equation}
    |-\rangle ~=~(0,1) ~~,~~
    |+\rangle  ~=~ (1,0)~~,
\end{equation}
and note further that $\langle\lambda d\lambda\rangle=dz$ and $\langle\tilde{\lambda}d\tilde{\lambda}\rangle=idx$. We can now rewrite \eqref{eq:samemod} as
\begin{equation}
    \frac{1}{2\pi i}\oint_{S^{1}} \langle\tilde{\lambda} d\tilde{\lambda}\rangle \, \frac{\langle\tilde{\lambda}+\rangle^{k-1/2}}{\langle\tilde{\lambda}-\rangle^{k+1/2}}\,\tilde{\mu}_{\alpha}(\tilde{\lambda}) ~=~ \frac{1}{2\pi i}\oint_{\mathbb{RP}^{1}}\langle\lambda d\lambda\rangle\,\frac{\langle\lambda+\rangle^{k-1/2}}{\langle\lambda-\rangle^{k+1/2}}\,\mu_{\alpha}(\lambda) ~~.
\end{equation}
In a global coordinate patch, we can assume the function $\mu(\lambda)$
can be analytically extended and has singularities only at $\langle\lambda+\rangle=0$
and $\langle\lambda-\rangle=0$ related to operator insertions. In
such case, we can deform the second contour $\lambda\to\tilde{\lambda}=\Lambda\lambda$
using the ${\rm SL}(2,\mathbb{C})$ map~\cite{Guevara:2023tnf}. 
We obtain
\begin{equation}
    \frac{1}{2\pi i}\oint_{S^{1}}\langle\tilde{\lambda}d\tilde{\lambda}\rangle\,\frac{\langle\tilde{\lambda}+\rangle^{k-1/2}}{\langle\tilde{\lambda}-\rangle^{k+1/2}}\,\tilde{\mu}_{\alpha}(\tilde{\lambda})~=~\frac{1}{2\pi i}\oint_{S^{1}}\langle\tilde{\lambda}d\tilde{\lambda}\rangle\,\frac{\langle\tilde{\lambda}+\rangle^{k-1/2}}{\langle\tilde{\lambda}-\rangle^{k+1/2}}\,\mu_{\alpha}(\tilde{\lambda}) ~~.
\end{equation}
This equivalence holds for all $k\in\mathbb{Z}$. Since $\mu_{\alpha}$ and
$\tilde{\mu}_{\alpha}$ now have the same moments along the circle they must
coincide on it. Hence $\tilde{\mu}_{\alpha}$ is determined by the unique analytic
continuation of $\mu_{\alpha}$. More importantly, this also provides a prescription
to evaluate the contour on $\mathbb{RP}^{1}$ via a compactification
to $S^{1}$.

It may be sometimes convenient to provide an explicit expression for the homogeneous
function $\mu_{\alpha}(\lambda)$. For a finite range such as \eqref{eq:eran} it is given by
\begin{equation}
    \mu_{\alpha}(\lambda) ~=~ \frac{W_{\alpha,\dot{\alpha}_{1}\cdots\dot{\alpha}_{N-1}}\, \lambda^{\dot{\alpha}_{1}}\cdots\lambda^{\dot{\alpha}_{N-1}}}{\langle\lambda+\rangle^{N/2}\langle\lambda-\rangle^{N/2}} ~~,
\end{equation}
where $W_{\alpha,\dot{\alpha}_{1}\cdots\dot{\alpha}_{N-1}}=W_{\alpha,(\dot{\alpha}_{1}\cdots\dot{\alpha}_{N-1})}$ is symmetric
in the dotted indices. Among other things, this shows explicitly that the weight of $\mu_{\a}$ is $h=1/2$ as a homogeneous function of $\lambda$. For $N=2$ we have $W_{\alpha,\dot{\alpha}}=x_{\alpha\dot{\alpha}}$, namely
\begin{equation}
\mu_{\alpha}(\lambda)~=~\frac{x_{\alpha\dot{\alpha}}\, \lambda^{\dot{\alpha}}}{\langle\lambda+\rangle \langle\lambda-\rangle }   ~~.
\end{equation}
By virtue of \eqref{eq:2params}, this is precisely the qubit field
\begin{equation}
    \mu_{\alpha}(z) ~=~ x_{\a +} ~+~ \frac{x_{\a -}}{z}
\end{equation}
that we introduced in \cite{Guevara:2023tnf}, and matches the weight $h=1/2$ model defined in~\cite{Adamo:2021bej}. In general,  $W_{\alpha,\dot{\alpha}_{1}\cdots\dot{\alpha}_{N-1}}$ matrix elements can be mapped to $\mu^{(k)}_{\a}$ via the formula 
\begin{equation}   
\binom{N-1}{\frac{N-1}{2}-k}\,W_{\alpha,(\underbrace{+\cdots+}_{\frac{N-1}{2}+k}\,\underbrace{-\cdots-}_{\frac{N-1}{2}-k})} ~=~ \mu_{\a}^{(k)} ~~.
\end{equation}

Finally let us discuss the role of ${\rm SL}(2,\mathbb{R})$ transformations on $\mu^{(k)}_{\a}$. Denote the transformation by $\mu^{(k)}_{\a}\to \mu'^{(k)}_{\a}$. We have 
\begin{equation}
     \mu'^{(k)}_{\a} ~=~ \frac{1}{2\pi i}\oint_{S^{1}}\langle\tilde{\lambda}d\tilde{\lambda}\rangle\,\frac{\langle\tilde{\lambda}+\rangle^{k-1/2}}{\langle\tilde{\lambda}-\rangle^{k+1/2}}\,\tilde{\mu}_{\a}(\tilde{\Sigma}\tilde{\lambda}) ~=~ \frac{1}{2\pi i}\oint_{\mathbb{RP}^{1}}\langle\lambda d\lambda\rangle\,\frac{\langle\lambda+\rangle^{k-1/2}}{\langle\lambda-\rangle^{k+1/2}}\,\mu_{\a}(\Sigma\lambda) ~~.
\end{equation}
Here $\tilde{\Sigma }$ and $\Sigma$ are ${\rm SL}(2,\mathbb{R}) $ matrices that preserve the contours $S^1$ and $\mathbb{RP}^1$ respectively. To understand how they are related, recall that $\tilde{\mu}_{\a}$ is the analytic continuation of $\mu_{\a}$ via the contour deformation $\lambda \to \tilde{\lambda }=\Lambda \lambda$. Hence
\begin{equation}
    \tilde{\Sigma} \tilde{\lambda}~=~ \Lambda \Sigma \lambda ~~,
\end{equation}
which yields
\begin{equation}
    \tilde{\Sigma} ~=~ \Lambda \Sigma \Lambda^{-1}~~.
\end{equation}

\section{\texorpdfstring{SU($N$) Representations}{SU(N) Representations}}\label{appen:SU(N)}

In this appendix, we briefly review the statement that the $\{{\mu}_{\pm}\}$ system with commutation relation 
\begin{equation}
    \big[{\mu}_{-},\,{\mu}_{+}\big] ~=~ \frac{2\pi i}{N}
\end{equation}
is a representation of ${\rm SU}(N)$, following~\cite{Hoppe:1988gk}. 
Indeed, the generator of SU($N$) ($N$ is even\,\footnote{For odd $N$, we choose $[{\mu}_{-},\,{\mu}_{+}] = \frac{4\pi i}{N}$ and $q=e^{\frac{4\pi i}{N}}$ has period $N$. Then the structure constant reduces to $\sin\left( \frac{2\pi}{N}[mn]\right)$~\cite{Hoppe:1988gk,FAIRLIE1989203}.}) can be represented as follows
\begin{equation}
    T_{m} ~=~ i\frac{N}{2\pi}\,e^{i[m{\mu}]} ~=~ i\frac{N}{2\pi}\,q^{m^+m^-/2}{g}_{-}^{m^-}\,{g}_{+}^{m^+} ~~,
    \label{equ:Tm}
\end{equation}
where 
\begin{equation}
    q~=~e^{\frac{2\pi i}{N}}~~,~~{g}_{\pm} ~=~ e^{i\,{\mu}_{\pm}}~~,
\end{equation}
and $m^{\pm}$ is exactly living in the stabilizer lattice, i.e. $(m^{-}, m^+)\in \mathbb{Z}_N\times \mathbb{Z}_N$. 
Hence there are $N^2$ of $T_m$ corresponding to $N^2$ generators of ${\rm U}(N)$ group. 
Excluding the origin of the lattice (i.e. $m^{\pm}=0$), there are $N^2-1$ of $T_m$, matching with the dimension of ${\rm SU}(N)$. 
Moreover, $T_m$ defined in (\ref{equ:Tm}) satisfies 
\begin{equation}
    T_mT_n ~=~ i\frac{N}{2\pi}\,q^{-[mn]/2}\,T_{m+n} ~~.
\end{equation}
Hence
\begin{equation}
    \Big[ T_m,\, T_n \Big] ~=~ i\frac{N}{2\pi}\,\left( e^{-\pi i[mn]/N} - e^{\pi i[mn]/N} \right)\,T_{m+n} ~=~ \frac{N}{\pi}\,\sin\left(\frac{\pi}{N}[mn]\right)\,T_{m+n} ~~.
\end{equation}

Finally, note that ${g}_{\pm}$ forms the Heisenberg group
\begin{equation}
    {g}_{+}{g}_{-} ~=~ q\,{g}_{-}{g}_{+}
\end{equation}
and can be represented by $N\times N$ unitary traceless matrices as follows
\begin{equation}
    {g}_+ ~=~ \begin{pmatrix}
    1 & & &\\
    & q & &\\
    &  & \ddots & \\
    & & & q^{N-1}
    \end{pmatrix}~~,~~
    {g}_- ~=~ \begin{pmatrix}
    0 & 1 &   &  \\
      & 0 & 1 &   \\
      &   & \ddots & \ddots \\
    1 &   &  &  0 
    \end{pmatrix} ~~.
\end{equation}

\section{Truncated Virasoro Algebra}\label{appen:dvir}

In section~\ref{sec:sym}, we have examined the continuous symmetry ${\rm Sp}(2N,\mathbb{R})$ that preserve (\ref{equ:mu-mode-exp}). In particular, we found subgroups that factorize into the left and right sectors and are mutually commuting ${\cal V}_{\rm left}\times {\cal V}_{\rm right}$. From the CFT perspective, this factorization corresponds to transforming $z$ and $\bz$ independently. Based on (\ref{equ:truncated-mode-exp}) where the phase space variables $\mu_{\a}^{(k)}$ for qudits emerge as modes of a conformal field, ${\cal V}_{\rm left}$ corresponds to solely transforming the mode index $k$ while leaving the spinor index $\a$ invariant. From section~\ref{sec:sym} we know this is ${\rm GL}(N)$. Indeed, this corresponds to rotations acting only on the mode indices, namely
\begin{equation}
     \mu_{\a}^{(k)}~\mapsto~ \sum_l\,(\L_{\a})^{k}{}_l\,\mu_{\a}^{(l)} ~~,~~
     \sum_l\,(\L_{+})^{k}{}_{-l}\, (\L_{-})^{-k'}{}_{l} ~=~ \d^{k,k'} ~~. 
    \label{equ:mode-index-transf}
\end{equation}
Note that when $N=2$ this reduces to the transformation in (2.22) of \cite{Guevara:2023tnf} by using (4.6) there.  
If $\mu_{\pm}$ were to be Hermitian conjugates, so would $\Lambda_{\pm}$, and this would represent $U(N)$. In our `Kleinian' setup however, these are two real matrices and we can take $\Lambda_{+}$ to be an arbitrary generator of ${\rm GL}(N,\mathbb{R})$, with $\Lambda_-$ being essentially its inverse.

Notice that $\Lambda_{\pm}$ has a spinor index and hence transforms under ${\rm SL}(2,\mathbb{R})_{\textrm{right}}$, which precludes us from factorizing left and right symmetry. However, this is resolved if we consider the intersection of the operation (\ref{equ:mode-index-transf}) with the diagonal subgroup $\Lambda_{+}=\Lambda_{-}$, and we obtain
\begin{equation}
    \sum_l\,\L^{k}{}_{-l}\, \L^{-k'}{}_{l} ~=~ \d^{k,k'} ~~,  
    \label{equ:SON}
\end{equation}
which is the ${\rm SO}(N)_{\rm left}$ subgroup of ${\rm GL}(N)$. Remarkably, this ${\rm SO}(N)_{\rm left}$ group contains the `truncated' version of the Virasoro algebra $\widehat{{\rm Vir}}_{\rm left}$, which is the theme of this appendix.

Indeed there is a subset of elements in ${\rm SO}(N)$ group that becomes Virasoro infinitesimally. This can be seen as follows. 
First, we note that any element in ${\rm SO}(N)$ can be decomposed in terms of $N(N-1)/2$ independent rotations. 
Among them, we consider the following subset that contains $(2N-3)$ independent rotations
\begin{equation}
    \begin{split}
    %%%%%%%%%%%%%%%%%%%%%%%%%%%%%%%
     \left\{ \scalemath{0.9}{  \begin{pmatrix}
  \begin{matrix}
  \cos\frac{\theta}{2} & \sin\frac{\theta}{2} \\
  -\sin\frac{\theta}{2} & \cos\frac{\theta}{2}
  \end{matrix}
  & \rvline & \mathbf{0} \\
\hline
  \mathbf{0} & \rvline &
  \mathbf{1}
\end{pmatrix}},  
%%%%%%%%%%%%%%%%%%%%%%%%%%%%%%%
\scalemath{0.8}{
\begin{pmatrix}
  \begin{matrix}
  \cos\theta &  0 & \sin\theta \\
  0 & 1 & 0\\
  - \sin\theta & 0 & \cos\theta
  \end{matrix}
  & \rvline & \mathbf{0} \\
\hline
  \mathbf{0} & \rvline &
  \mathbf{1}
\end{pmatrix}}\,,\cdots,\, 
%%%%%%%%%%%%%%%%%%%%%%%%%%%%%%%
\scalemath{0.7}{
    \begin{pmatrix}
    \cos\frac{(n-1)\theta}{2} &   &  & & & \sin\frac{(n-1)\theta}{2} \\
     & \cos\frac{(n-3)\theta}{2} &  & ~ & \sin\frac{(n-3)\theta}{2} & ~ \\
     &  & \ddots & \reflectbox{$\ddots$} & ~ & \\
    ~ & \sin\frac{(3-n)\theta}{2} & ~ & ~ & \cos\frac{(n-3)\theta}{2} & ~ \\
    \sin\frac{(1-n)\theta}{2} & ~ & ~ & ~ & ~&  \cos\frac{(n-1)\theta}{2} \\
    \end{pmatrix}}~, \right. & \\
%%%%%%%%%%%%%%%%%%%%%%%%%%%%%%%
%%%%%%%%%%%%%%%%%%%%%%%%%%%%%%%
   \left. \cdots~,~ \scalemath{0.8}{
   \begin{pmatrix}
    \mathbf{1}
  & \rvline & \mathbf{0} \\
\hline
  \mathbf{0} & \rvline &
\begin{matrix}
  \cos\theta &  0 & \sin\theta \\
  0 & 1 & 0\\
  - \sin\theta & 0 & \cos\theta
  \end{matrix}
\end{pmatrix}
   }
%%%%%%%%%%%%%%%%%%%%%%%%%%%%%%%
~,~ \scalemath{0.9}{\begin{pmatrix}  
   \mathbf{1} & \rvline & \mathbf{0} \\
\hline
  \mathbf{0} & \rvline &
 \begin{matrix}
  \cos\frac{\theta}{2} & \sin\frac{\theta}{2} \\
  -\sin\frac{\theta}{2} & \cos\frac{\theta}{2}
  \end{matrix}
\end{pmatrix}} \right\} ~~. &
    \end{split} 
    \label{equ:finiteVir-matrix}
\end{equation}
Expanding around $\theta=0$, their infinitesimal transformations take the same form as the actions of ${\rm Vir}_{\rm left}$ generators on the mode as follows 
\begin{equation}
    \big[ L_m,\, \mu^{(k)}_{\a} \big] ~=~ \left( -\frac{m}{2} - k \right)\,\mu^{(m+k)}_{\a} ~~,
    \label{equ:transf-Ln}
\end{equation}
where $m\in\left[2-N,N-2\right]$ and $\mu^{(k)}_{\a}=0$ for $|k|\ge N/2$, followed by (\ref{equ:truncated-mode-exp}).

Although $k$ is truncated by construction, if we forget about (\ref{equ:finiteVir-matrix}), which is a representation of (\ref{equ:transf-Ln}), $m$ is not truncated a priori. Therefore one should continue the $m$ in (\ref{equ:transf-Ln}) to take values in all integers (i.e. the full Virasoro) and ask do they still preserve the commutation relation (\ref{equ:mu-mode-exp}). 
The answer is yes, however, only $2N-3$ of $L_m$'s act nontrivially, which are $\{ L_{2-N}, L_{3-N}, \cdots, L_{N-2} \}$. Any $L_m$ with $|m|>N-1$ transforms every mode to the one outside the truncation, therefore their actions are all trivial. In terms of $N\times N$ matrices, all entries in these matrices are zero. 
For $L_{\pm (N-1)}$, the only action that could be nontrivial is on the mode $\mu_{\a}^{(\mp\frac{N-1}{2})}$. However, the coefficient on the RHS of (\ref{equ:transf-Ln}) is exactly zero, so they are trivial as well. 
When $N\to\infty$, namely $k$ runs from $-\infty$ to $+\infty$, clearly $\widehat{{\rm Vir}}_{\rm left}$ becomes the actual ${\rm Vir}_{\rm left}$.

\subsection{From Truncated to Discrete action on the circle}\label{appen:discrete-diff}

Since the $k$ truncation naturally emerges in (\ref{equ:truncated-mode-exp}), this brings us to the question: what is the implication of the $\widehat{{\rm Vir}}_{\rm left}$ on the field and can it be interpreted as a discrete version of the diffeomorphism, from which the diffeomorphism on a circle can be restored by taking the large $N$ limit?

The idea is that so far we have identified the Virasoro symmetry from its action on the modes (\ref{equ:transf-Ln}). A more natural way to define conformal symmetry is starting from the transformation in the $z$ plane. Therefore, for the case studied in this work, we would like to define the truncated Virasoro from a discrete version of diffeomorphism. Then the question becomes: 

\begin{center}
\parbox{0.8\textwidth}{
\textit{What does a ``diffeomorphism on the circle" mean when there is only a discrete number of points on the circle?}}
\end{center}

Apparently, we cannot take the Lie derivative anymore. To answer this question, let's start with a more straightforward one -- the first question we raised at the beginning of this section.

Recall that with finite $N$, our field 
\begin{equation}
    \mu_{\a}(z_j) ~=~ \sum_{k=-\frac{N-1}{2}}^{\frac{N-1}{2}}\,\mu_{\a}^{(k)}\,z_j^{k} 
    ~~,~~ j~=~0,\cdots,N-1
    \label{equ:muz}
\end{equation}
only takes values at discrete positions $z_j=e^{\frac{2\pi i}{N}j}$ rather than being a function of a continuous variable. 
Moreover, the mode can be extracted as follows
\begin{equation}
    \mu_{\a}^{(k)} ~=~ \frac{1}{N}\,\sum_{j=0}^{N-1}\,\mu_{\a}(z_j)\,z_j^{-k} ~~.
    \label{equ:mu-extraction}
\end{equation}
Given (\ref{equ:transf-Ln}), the field (with generic weight $h$) transforms under $L_m$ as follows
\begin{equation}
    \begin{split}
        \mathcal{L}_{m}\,\mu_{\a}(z_i)~=&~ \sum_{k=-\frac{N-1}{2}}^{\frac{N-1}{2}}\,\big[m(h-1)-k\big]\,\mu^{(k+m)}_{\a}\,z_{i}^{k}
        ~=~ \sum_{k=-\frac{N-1}{2}+m}^{\frac{N-1}{2}+m}\,\big[mh-k\big]\,\mu^{(k)}_{\a}\,z_{i}^{k-m}\\
        ~=&~ z_i^{-m}\,\big[ mh+\mathcal{L}_{0}\big]\,\sum_{k=-\frac{N-1}{2}+m}^{\frac{N-1}{2}+m}\,\mu^{(k)}_{\a}\,z_{i}^{k}~~,
        \label{equ:Lmdiscrete}
    \end{split}
\end{equation}
where we keep in mind that $\mu^{(k)}_{\a}$ is vanishing if $|k|\ge N/2$. 
In the continuum limit ($N\to\infty$), $z_{i}\to z\in\mathbb{R}$, we have 
\begin{equation}\label{equ:L0mu(z)}
    \mathcal{L}_{0}\,\mu_{\a}(z_i) ~=~ -\,\sum_{k=-\frac{N-1}{2}}^{\frac{N-1}{2}}\,k\,\mu^{(k)}_{\a}\,z_{i}^{k} ~~\overset{N\to\infty}{\longrightarrow}~~ \mathcal{L}_{0}~=~-z\partial_{z} ~~.
\end{equation}
In such case 
\begin{equation}
    \mathcal{L}_{m}\,\mu_{\a}(z) ~=~ z^{-m}\left(mh-z\partial_{z}\right)\mu(z)
\end{equation}
is the standard Lie derivative for a weight $h$ field. Therefore, we regard (\ref{equ:Lmdiscrete}) as the discrete version of the Lie derivative. 

From (\ref{equ:L0mu(z)}) our main observation is that ${\cal L}_0$ maps $\mu_{\a}(z_i)$ to a linear combination of the basis $\{\mu_{\a}(z_j)\}$ as follows
\begin{equation}
    \mathcal{L}_{0}\,\mu_{\a}(z_i) ~=~ \sum_{j=0}^{N-1}\,K_{i-j}\,\mu_{\a}(z_j)~~,
\end{equation}
where using (\ref{equ:mu-extraction}) we have 
\begin{equation}
    K_{i-j} ~=~ \frac{1}{N}\,\sum_{k=-\frac{N-1}{2}}^{\frac{N-1}{2}}\,k\,e^{\frac{2\pi i\,k}{N}(i-j)} ~=~ \begin{cases}
        0 \qquad\qquad~ i=j\\
        \frac{1}{N}\,\frac{(-1)^{i-j}}{2i\,\sin\frac{(i-j)\pi}{N}}~~i\ne j
    \end{cases} ~~.
\end{equation}
It is also straightforward to check that this mapping is symplectic. Namely, it preserves (\ref{eq:rr1p}). The canonical generator is simply
\begin{equation}
\mathcal{L}_{0} ~=~ -\frac{1}{i\tau}\,\sum_{i,j}\, :\mu_{+}(z_i) \mu_{-}(z_j):\, K_{i-j} ~~.
\end{equation}
Therefore, we can define the discrete diffeomorphism in the following way:

\begin{center}
\parbox{0.8\textwidth}{
\textit{A discrete ``diffeomorphism on the circle" is a symplectic map taking the field at discrete points $\mu_{\a}(z_i)$ to a linear combination of them.}}
\end{center}

We close this section by showing the recovery of the diffeomorphism from examining the finite transformation at $N\to\infty$. 
The finite transformation generated by $L_m$ can be obtained by exponentiation of (\ref{equ:transf-Ln}) 
and reads 
\begin{equation}
    \mu_{\a}^{(k)} ~\mapsto~ (\Lambda_m)^k{}_l\,\mu_{\a}^{(l)} ~~,~~ (\Lambda_m)^k{}_l 
        ~=~ \sum_{p=0}^{\infty} \frac{(-1)^p}{p!}\,\prod_{r=0}^{p-1}\left(\frac{m}{2}+rm+k\right)\,\d^{pm+k}_l ~~
        \label{equ:Lambda-Vir}
\end{equation}
with the convention of truncating modes $\mu_{\alpha}^{(l)}$ with $|l|\ge N/2$.
The transformation on the field (\ref{equ:muz}) becomes
\begin{equation}
    \begin{split}
        \mu_{\a}'(z) ~=&~ \sum_{k=-\frac{N-1}{2}}^{\frac{N-1}{2}}\,\mu_{\a}'^{(k)}\,z^{k}
        ~=~\sum_{k=-\frac{N-1}{2}}^{\frac{N-1}{2}}\left(\sum_{p=0}^{\infty}\frac{(-1)^{p}}{p!}\prod_{r=0}^{p-1}\left(\frac{m}{2}+rm+k\right)\mu_{\alpha}^{(k+pm)}\right)z^{k}\\
        ~\overset{N\to\infty}{=}&~\sum_{k=-\infty}^{\infty}\mu_{\alpha}^{(k)}\left(\sum_{p=0}^{\infty}\frac{(-1)^{p}}{p!}\prod_{r=0}^{p-1}\left(\frac{m}{2}+rm+k-pm\right)z^{k-pm}\right)\\
        ~=&~ (1-mz^{-m})^{-\frac{1}{2}}\,\sum_{k=-\infty}^{\infty}\mu_{\alpha}^{(k)}\,z^k\,(1-mz^{-m})^{\frac{k}{m}} ~=~ \left(\frac{d\log z'}{d\log z}\right)^{\frac{1}{2}}\,\sum_{k=-\infty}^{\infty}\mu_{\alpha}^{(k)}\,z'^{k}~~,
    \end{split}
\end{equation}
where $z'=z(1-mz^{-m})^{1/m}$ is a finite diffeomorphism and this is indeed the transformation of a weight $h=1/2$ under diffeomorphism on the circle.

\bibliographystyle{utphys}
\bibliography{reference}

\end{document}